\documentclass[useAMS,twocolumn]{aastex631}

\usepackage{amsmath,amssymb}
\usepackage{natbib}
\usepackage{xspace}
\usepackage{ulem}

\newcommand{\simgt}{\lower.5ex\hbox{$\; \buildrel > \over \sim \;$}}
\newcommand{\simlt}{\lower.5ex\hbox{$\; \buildrel < \over \sim \;$}}

\newcommand{\myemail}{masato.kobayashi@nao.ac.jp, mskobayashi@astr.tohoku.ac.jp, masato.kobayashi@nagoya-u.jp}

\newcommand{\m}{${\rm M}_{\rm 200b}$ }

\newcommand{\eg}{e.g.\xspace}
\newcommand{\cf}{c.f.\xspace}
\newcommand{\ie}{i.e.\xspace}
\newcommand{\etc}{etc.}

\newcommand{\msun}{\mbox{${\rm M_{\odot}}$}\xspace}

\newcommand{\kms}{{\rm km \, s^{-1}}}

\newcommand{\cmkk}{\mbox{${\rm cm^{-3}}$}}

\newcommand{\myCooL}{\mathcal{L}}

\newcommand{\myboltz}{k_{\rm B}^{}}
\newcommand{\mycs}{C_{\rm s}^{}}

\newcommand{\myss}{\sigma_{\rm s}^{}}
\newcommand{\myssim}{\sigma_{\rm s,sim}}
\newcommand{\mysiso}{\sigma_{\rm s,theory}}

\newcommand{\unitk}{\mathbf{\hat{k}}}

\received{--}
\revised{--}
\accepted{--}
\submitjournal{ApJ}

\begin{document}

%-------- TITLE  ---------------------

    \title{NATURE OF SUPERSONIC TURBULENCE 
    AND DENSITY DISTRIBUTION FUNCTION \\
    IN THE MULTIPHASE INTERSTELLAR MEDIUM}
    \shorttitle{SOLENOIDAL/COMPRESSIVE MODES DURING CLOUD FORMATION}
    \shortauthors{Masato I.N. Kobayashi}

%-------- AUTHORS  ---------------------

    \correspondingauthor{Masato I.N. Kobayashi}
    \email{\myemail}
    \author[0000-0003-3990-1204]{Masato I.N. Kobayashi}
    \affiliation{Division of Science, National Astronomical Observatory of Japan, 2-21-1 Osawa, Mitaka, Tokyo 181-8588, Japan }\\
    \affiliation{Astronomical Institute, Graduate School of Science, Tohoku University, Aoba, Sendai, Miyagi 980-8578, Japan}\\
    \author{Tsuyoshi Inoue}
    \affiliation{Division of Particle and Astrophysical Science, Graduate School of Science, Nagoya University, Aichi 464-8602, Japan}\\
    \affiliation{Department of Physics, Konan University, Okamoto 8-9-1, Kobe, Japan}\\
    \author[0000-0001-8105-8113]{Kengo Tomida}
    \affiliation{Astronomical Institute, Graduate School of Science, Tohoku University, Aoba, Sendai, Miyagi 980-8578, Japan}\\
    \author[0000-0002-2707-7548]{Kazunari Iwasaki}
    \affiliation{Center for Computational Astrophysics, National Astronomical Observatory of Japan, Mitaka, Tokyo 181-8588, Japan}\\
    \author{Hiroki Nakatsugawa}
    \affiliation{Division of Particle and Astrophysical Science, Graduate School of Science, Nagoya University, Aichi 464-8602, Japan}
   
%-------- ABSTRACT  ---------------------

\begin{abstract}
Supersonic flows in the interstellar medium (ISM) are believed to be a key driver 
of the molecular cloud formation and evolution. Among molecular clouds' properties, 
the ratio between the solenoidal and compressive modes of turbulence plays important roles
in determining the star formation efficiency. We use numerical simulations of 
supersonic converging flows of the warm neutral medium (WNM) resolving the thermal instability 
to calculate the early phase of molecular cloud formation, and investigate 
the turbulence structure and the density probability distribution function (density PDF) 
of the multiphase ISM. 
We find that both the solenoidal and compressive modes have their power spectrum 
similar to the Kolmogorov spectrum. The solenoidal (compressive) modes account for $\gtrsim 80$ \% 
($\lesssim 20$ \%) of the total turbulence power. 
When we consider both the cold neutral medium (CNM) and the thermally unstable neutral medium (UNM) 
up to T $\lesssim 400$ K, the density PDF follows the log-normal distribution whose width 
$\myss$ is well explained by the known relation from the isothermal turbulence 
as $\myss = \ln(1+b^2 \mathcal{M}^2)$ (where $b$ is the parameter representing 
the turbulence mode ratio and $\mathcal{M}$ is the turbulent Mach number). 
The density PDF of the CNM component alone (T $\leq 50$ K), however, exhibits 
a narrower $\myss$ by a factor of $\sim 2$. These results suggest that 
observational estimations of $b$ based on the CNM density PDF requires 
the internal turbulence within each CNM clump but not the inter-clump relative velocity, 
the latter of which is instead powered by the WNM/UNM turbulence.

\end{abstract}

\keywords{Interstellar medium, Warm neutral medium, Cold neutral medium, Interstellar dynamics}

%%%%%%%%%%%%%%%% 
% Introduction %
%%%%%%%%%%%%%%%% 
\section{Introduction}
\label{sec:intro}
Molecular clouds are progenitor of stars \citep{Kennicutt2012},
and their physical states set the
initial condition to determine the star formation rate/efficiency.
In particular, 
the density structure is important
where the densest volume of molecular clouds eventually collapse gravitationally
to form stars \citep{Krumholz2005}.

One of the key statistics is 
the (column) density probability distribution function (PDF).
Based on analytic models and one-dimensional simulations,
\cite{Passot1998} show that
isothermal turbulence produces 
a log-normal density PDF,
which is a compilation of density enhancement by multiple shocks.
Other analytic and numerical studies 
\citep[\eg,][]{Padoan1997,Nordlund1999} also find such a log-normal
density PDF and discuss its relation to the functional form of the stellar initial mass function 
(stellar IMF).
Two- and three-dimensional simulations later suggest that
the log-normal density PDF ubiquitously exists
under the driven turbulence 
\citep[\eg,][]{Federrath2008},
where the tails can deviate from the log-normal PDF 
due to non-isothermality \citep{Passot1998,Scalo1998},
self-gravity \citep{Klessen2000},
and magnetization effects \citep{Li2008}
\citep[see also the review by][]{Elmegreen2004}.
Molecular clouds in the Milky Way galaxy indeed exhibit log-normal PDFs
of (column) densities at low $A_{\rm v}$ regime, 
followed with a power-law tail at the high density end 
(\citealt[][]{Lombardi2011,Schneider2013,AlvesdeOliveira2014,Schneider2016};
but see also \citealt[][]{Lombardi2015}).

There have been theoretical studies
to estimate how the star formation rate varies
based on the formation of the densest parts of such density PDFs within molecular clouds,
by assuming some star formation efficiency (SFE) per free-fall time
in individual molecular cloud cores
\citep[\eg,][]{Krumholz2005,Federrath2008,Padoan2011}.
In particular, the suite of isothermal simulations with driven turbulence 
show that the turbulence mode ratio between solenoidal and compressive modes
as well as the turbulent Mach number are important parameters
that control the width of the density PDF \citep[\eg,][]{Federrath2008,Federrath2010}.
For example, the star formation rate varies by a factor to one order of magnitude 
depending on the turbulence mode ratio \citep{Federrath2012}.
This suggests that different density PDFs and the turbulence mode ratios
result in various star formation rate
of star-forming clouds
even under a fixed turbulence power spectrum.
Meanwhile, detailed simulations of radiative feedback from massive stars 
suggest that the SFE depends strongly on the initial column density of the cloud
\citep[\eg,][]{KimJG2018,Fukushima2020c,Fukushima2021a},
and the turbulent structure likely
impacts the stellar initial mass function as well
\citep{Padoan2002}.
Therefore, it is important to reveal the turbulence structure 
and resultant density structures achieved in molecular clouds under realistic conditions 
of molecular cloud formation and evolution.

Supersonic flows in the interstellar medium (ISM) 
are one of the key driving sources of the turbulence.
In galactic disk regions of star-forming galaxies (like the Milky Way galaxy), the typical interval 
between the passages of successive supersonic flows at any volume of the ISM is 
about $1$ Myr \citep{McKee1977}, where those flows originate in
supernovae, expanding H{\sc ii} regions, galactic spirals \etc.
The typical molecular cloud lifetime is also believed to be about a few 10 Myr, 
(\citealt{Kawamura2009}; see also \citealt{Meidt2015});
where molecular clouds finally evaporate
within 10 Myr due to the stellar feedback once massive stars form
\citep{Hosokawa2006a,Kruijssen2019,Fukushima2020c,KimJG2021}.
The interaction between molecular clouds and supersonic flows
is naturally expected to occur multiple times during the cloud lifetime.
\cite{Matsumoto2015} perform simulations 
of an isothermal ISM 
to investigate such interactions
and show that
the turbulent velocity 
within molecular clouds
does not significantly differ
between with and without the colliding flows
if averaged over $1$ -- $3$ free-fall time 
of the cloud 
\citep[see also][]{Seifried2018},
but the turbulent mode ratio as well as the density PDF 
does differ, resulting in a rapid star formation
due to the colliding flows.
\cite{Padoan2016} perform simulations of multiple supernovae on a 100 pc scale
and find that the expansion of multiple supernova remnants 
leads to drive the solenoidal modes stronger than the compressive modes
by a factor of about 5. 

Supersonic flows drive not only the turbulence 
but also the evolution of the thermal sates
by triggering
the phase transition 
from the warm neutral medium (WNM)
to the cold neutral medium (CNM)
through the thermal instability
\citep[\eg,][]{Field1965,Balbus1986}.
This transition from the WNM to the CNM 
increases the density,
and is believed to be an important first step
of molecular cloud formation
\citep{Koyama2002,Audit2005}.
Therefore, simultaneous investigations of both the turbulent structure
and thermal states  
provide the comprehensive understanding of the evolution of 
the density structure within molecular clouds
and resultant SFR\@.
However, such studies are still limited 
and focus on the later stages of molecular cloud evolution
where molecular clouds are massive enough to be self-gravitating
\cite[\eg,][]{Kortgen2017}.

As a complementary study to understand the molecular cloud evolution 
comprehensively from its early phase along with the thermal evolution,
we perform hydrodynamics
simulations of supersonic converging WNM flows,
by resolving the thermal instability 
and the formation of the multi-phase ISM as
a precursor of molecular clouds.
We investigate 
the density PDF and turbulence structure 
in this multi-phase ISM\@.
We measure the turbulence power spectrum, velocity structure functions,
CNM clump size and mass spectra,
the density PDF as a function of temperature,
and the turbulence mode ratio on 
$0.01$ -- $10$ pc scales,
based on which we show 
the range that the isothermal turbulence framework 
explains the density PDF in the multi-phase ISM.

The rest of this article is organized as follows. 
In Section~\ref{sec:method}, we 
explain 
the method of our simulations and analyses.
In Section~\ref{sec:key}, we explain the basic 
equations that describe the density PDF and the known
relation in the isothermal turbulence theory framework,
as well as the definition of the solenoidal/compressive modes.
We present the main results in Section~\ref{sec:allvol}
by using the data from the entire simulation domain,
whereas we focus on the results from nine sub-volumes
in Section~\ref{sec:subvol}.
In Section~\ref{sec:discussions},
we discuss the implications from our results
and future prospects.
We summarize our results in Section~\ref{sec:concl},
followed with Appendices 
explaining 
the convergence with respect to the spatial resolution
and additional information 
of the technical details in our analyses.

%%%%%%%%%% 
% Method %
%%%%%%%%%% 
\section{Method}
\label{sec:method}
\begin{figure*}
    \hspace{1.0cm}{\large (a) $\log (n\, [\mathrm{cm}^{-3}])$} 
    \hspace{5.5cm} {\large (b) $\log (T\, [\mathrm{K}])$ }  \\
    \begin{minipage}[t]{0.47\textwidth}
        \centering{\includegraphics[scale=0.57]{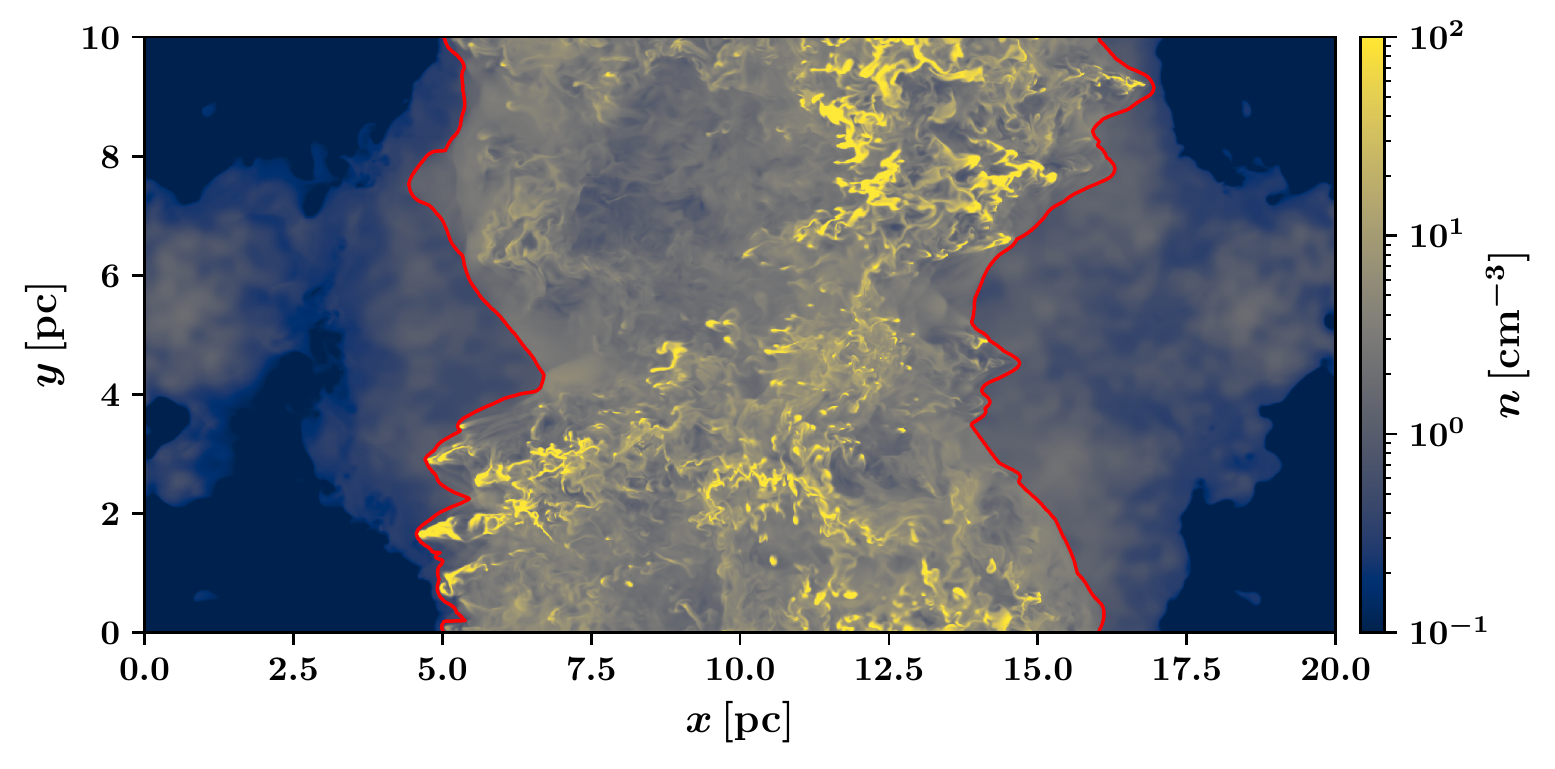}}
    \end{minipage}
    \hspace{0.5cm}
    \begin{minipage}[t]{0.47\textwidth}
        \centering{\includegraphics[scale=0.57]{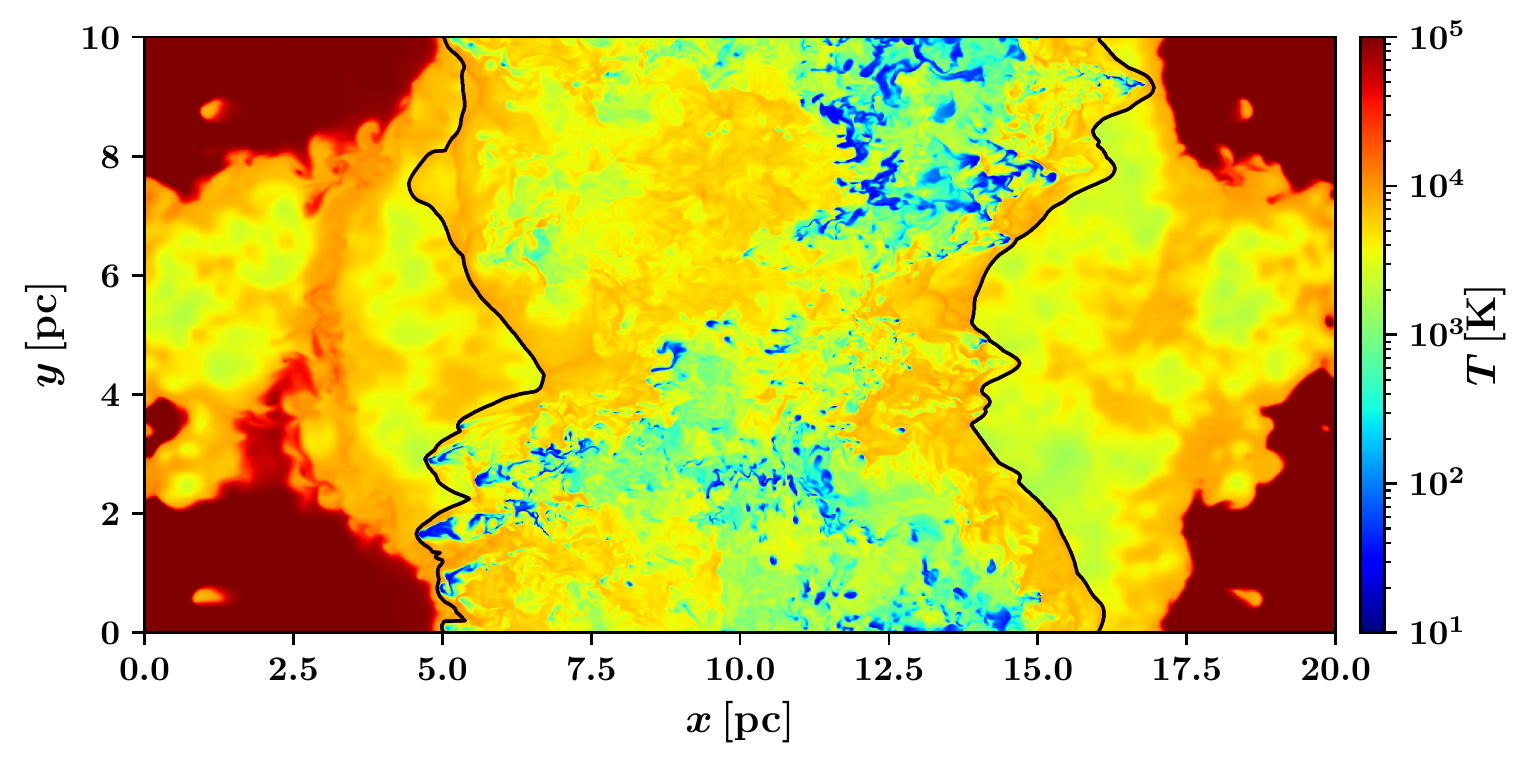}}
    \end{minipage}\\

    \caption{Panel (a): The density map from our three-dimensional simulation at 2.5 Myr.
             The central volume sandwiched by two shock fronts (thin red curves)
             is the progenitor of molecular clouds, within which the multi-phase ISM forms.
             Panel (b): Same as (a) but the temperature map with the shock positions
             shown as thin black curves.
             }
    \label{fig:rhoT}
\end{figure*}

\subsection{Basic Equations and Setups}
\label{subsec:code}
We utilize the simulation code and setups from our previous study \citep{Kobayashi2020}.
Our code is based on the hydrodynamics part from the magneto-hydrodynamics code originally developed in
\cite{Inoue2008}. 
This code employs the second-order Godunov scheme \citep{vanLeer1979}
and explicitly time-integrates the heating and cooling 
with the second order accuracy.
We solve the following basic equations: 
\begin{eqnarray}
    &&\frac{\partial \rho}{\partial t} + \nabla_{i}^{} (\rho v_{i}) = 0 \,, \label{eq:CoM}\\
    &&\frac{\partial (\rho v_{i})}{\partial t} 
       + \nabla_{j} (P\delta_{ij} + \rho v_{j}v_{i}) = 0 \,, \label{eq:EoM}\\
    &&\frac{\partial e}{\partial t} + \nabla_{i} \left( (e + P)v_{i} \right) 
       = \nabla_{i}(\kappa(T)\nabla_{i} T) -\rho \myCooL(T) \label{eq:EE}\,.
\end{eqnarray}
$\rho$ is the mass density, $v$ represents the velocity,
$P$ represents the thermal pressure, $T$ is the temperature, and 
$\nabla_{i} = \partial/\partial x_i$,
where $x_i$ spans $x, y,$ and $z$.
$\delta_{ij}$ is the identity matrix.
We calculate 
the total energy density,
$e$, as $e = P/(\gamma-1) + \rho v^2/2$
where $\gamma=5/3$ is the ratio of the specific heat. 
We implement the thermal conductivity, $\kappa$, 
as $\kappa(T) = 2.5 \times 10^3 \, T^{0.5} \, 
\mathrm{erg \, cm^{-1} \, s^{-1} \, K^{-1}}$,
representing collisions
between hydrogen atoms \citep{Parker1953}.

The thermal state of the gas is controlled by the balance between the cooling and heating, 
which is implemented in the net-cooling rate per mass $\myCooL(T)$.
We 
use the functional form of $\myCooL(T)$ that we develop in our previous study
\citep{Kobayashi2020}. This is the combination of the results from
\cite{Koyama2002} in $T \leq 14,577$ K and \cite{Cox1969} and \cite{Dalgarno1972} in $T > 14,577$ K
by considering the cooling rate due to
Ly$\alpha$, C$_{\rm II}$, He, C, O, N, Ne, Si, Fe, and Mg lines
with the photo-electric heating,
which corresponds to the typical ISM condition in the Milky Way galaxy.
In general, the shock-heated WNM experiences the thermally unstable phase
($(\partial (\mathcal{L}/T)/\partial T)_P <0$; \citealt[see \eg,][]{Balbus1986,Balbus1995}),
leading to the formation of the CNM clumps.
The multi-phase that appear in our simulations 
roughly correspond to 
the WNM as $T\geq5000$ K, the thermally unstable neutral medium (UNM) as $100 \leq T <5000$ K, and 
the CNM as $T<100$ K based on $(\partial (\mathcal{L}/T)/\partial T)_P$.
We refer the readers to \cite{Kobayashi2020} for the exact form of this $\myCooL(T)$.

We prepare a three-dimensional simulation domain with its size of 
$L_{x,y,z}=20,10,10$ pc,
and calculate the head-on collision of supersonic WNM converging flows
continuously injected through the two $x$ boundaries over 3 Myr.
We employ the periodic boundary condition on both $y$ and $z$ boundaries.
This flow creates a shock-compressed layer sandwiched by two shock fronts
at the box center $x\simeq L_{x}/2$ where the two flows collide.
The flow velocity is fixed as 20 km s$^{-1}$ representing 
common situations in galactic disks (\eg, 
the late phase of supernova remnant expansion, H {\sc ii} region expansion,
and normal shocks due to galactic spirals).
The injected WNM flow is thermally stable
with the mean number density $n_0=0.57 \,\cmkk$
and pressure $P_0/\myboltz= 3500 \, \mathrm{K \, cm^{-3}}$.
This corresponds to the mean temperature of $6141$ K,
the sound speed of $\mycs=8.16\,\kms$, and
the ram pressure of the converging flow as $P/\myboltz = 2.6 \times 10^4 \, \mathrm{K\,cm^{-3}}$.

The density of the injected WNM flow has a fluctuation 
with the Kolmogorov power spectrum $P_{\rho}(k)\propto k^{-11/3}$ 
\citep{Kolmogorov1941,Armstrong1995}.
The highest frequency is $k/2\pi=12.5$ pc$^{-1}$
(\ie, the wavelength of 0.08 pc).
The amplitude is set such that the mean dispersion 
is $\sqrt{\langle \delta n_0^2 \rangle}/n_0 = 1$
as the most realistic parameter in our calculation 
to describe the density structure in star-forming regions
\citep{Kobayashi2020}.
The interaction between this density inhomogeneity and shock fronts
induces turbulence in the shock-compressed layer \citep[\eg,][]{Inoue2012,CarrollNellenback2014}.

We employ uniform spatial resolution of 0.01 pc
in fiducial analyses, with additional resolutions
of 0.02 pc, 0.04 pc, 0.08 pc 
for comparision (Appendix~\ref{sec:ndiff}).
This 0.01 pc is motivated by our previous systematic study
over a wide parameter range of $\sqrt{\langle \delta n_0^2 \rangle}/n_0$
by \cite{Kobayashi2020}. In \cite{Kobayashi2020}, we show that,
with the thermal instability,
the spatial resolution with $\lesssim\, 0.02$ pc is required 
to fully resolve the cooling length of the UNM
and have the convergence in the mean density of the shock-compressed layer.
This spatial resolution resolves the highest frequency of the injected 
density fluctuation with at least eight cells.
In Appendix~\ref{sec:ndiff}, we also show that CNM clump properties
have their convergence with the 0.01 pc resolution compared with 0.02 pc
and 0.04 pc (such as mean density, size and mass spectra).

The calculation results are sampled every 0.1 Myr,
and we identify the shock front position 
by $P>1.3P_0$ to define 
the shock-compressed layer in each sampled time.
Figure~\ref{fig:rhoT} shows an example of the two-dimensional slices of the density distribution (Panel (a)) and the temperature distribution (Panel (b)) at 2.5 Myr. 
The central region sandwiched by the two shock fronts is the progenitor of molecular clouds,
within which the multi-phase ISM forms through the thermal instability.
The effective typical cooling length 
peaks at $\sim 1$ pc for the UNM and at $0.02$--$0.2$ pc for the CNM 
(see Panel (a) of Figure~\ref{fig:CNMhisto_ndiff}).
Therefore, through the growth of the thermal instability, the CNM structures develop  
on sub-pc scales as seen as clumpy structures on $\sim 0.1$ pc in Figure~\ref{fig:rhoT}.

\subsection{The Decomposition into the Solenoidal and Compressive Modes
and Our Choice of Analyzed Volume}
\label{subsec:decomp}
We apply the Fourier transformation to the shock-compressed layer 
to obtain the Fourier component of the velocity field
by using the Fast Fourier Transform in the West
(FFTW 3.3; \citealt{FFTW2005}),
and decompose this into the solenoidal modes and compressive modes
(see Section~\ref{sec:key}).
We perform this decomposition both on the entire volume of the shock-compressed layer 
and on nine sub-volumes of $(2.56$ pc$)^3$ size within the shock-compressed layer.
We summarize the technical details involved in this decomposition
in Appendix~\ref{sec:appA}.
We employ five temperature thresholds to 
extract volumes with various thermal states
and investigate
the difference in the turbulent structure and density PDF 
between the WNM+UNM and CNM;
$T\leq 50$ K, $T\leq 200$ K, $T\leq400$ K, $T\leq600$, and $T\leq1000$ K.

%%%%%%%%%%%%%%%%%%%%%%%%%%%%%%%%%%%%%%%%
% Key Points in Isothermal Turbulence  %
%%%%%%%%%%%%%%%%%%%%%%%%%%%%%%%%%%%%%%%%
\section{Density PDF in Isothermal Turbulence}
\label{sec:key}
The log-normal density PDF is defined as 
\begin{equation}
    p_{\rm s}(s) = \frac{1}{\sqrt{2\pi \sigma_s^2}} \exp \left( -\frac{(s-s_0)^2}{2\sigma_s^2} \right) \,,
    \label{eq:PS}
\end{equation}
where $s$ is the natural logarithm of density.
In the framework of the isothermal turbulence, essentially three parameters determine 
the star formation rate by controlling the fraction of dense volume
that gravitationally collapse to form stars. One is the mean density,
another is the mean Mach number of turbulence,
and the other is the mode ratio between the 
solenoidal/compressive modes.
In the case of such isothermal turbulence,
it is known that the width of this log-normal density PDF
$\sigma_s$, is well described by the combination of 
the forcing parameter $b$ and the mean Mach number $\mathcal{M}$ as 
\begin{equation}
    \mysiso^2 = \ln \left( 1 + b^2 \mathcal{M}^2 \right) \,,
    \label{eq:ssiso}
\end{equation}
(see \eg, \citealt[]{Federrath2010}).
$b$ represents the compressibility defined by the compressive ratio $\chi$ \citep{Pan2016} as:
\begin{eqnarray}
    b    &=& \sqrt{\frac{\chi}{1+\chi}} \,, \\
    \chi &=& \frac{ \langle \mathbf{\tilde{v}}_{\rm comp}^2 \rangle }{ \langle \mathbf{\tilde{v}}_{\rm sol}^2 \rangle } \,.
\end{eqnarray}
Here $\mathbf{\tilde{v}}_{\rm comp}$ and $\mathbf{\tilde{v}}_{\rm sol}$ mean 
the compressive and solenoidal modes of the turbulent velocity field, respectively.

Any velocity field 
is in principle able to be decomposed into 
the solenoidal (transverse) and compressive (longitudinal) modes
through the Helmholz decomposition as:
\begin{eqnarray}
    \mathbf{\tilde{v}}_{\rm sol}(\unitk)  &=&  \left( \unitk \times \mathbf{\tilde{v}} \right)  \times \unitk \\
    \mathbf{\tilde{v}}_{\rm comp}(\unitk) &=&  \left( \unitk \cdot  \mathbf{\tilde{v}} \right)         \unitk \,.
\end{eqnarray}
Here $\unitk$ is a unit wave vector and $\mathbf{\tilde{v}}$ is the Fourier component of the velocity field.
In the following analyses, we will measure 
$b$, $\mathcal{M}$, and $\sigma_s$ from our simulations,
and evaluate $\mysiso$ based on Equation~\ref{eq:ssiso}
to investigate the difference between $\sigma_s$
and $\mysiso$ in various thermal states.

%%%%%%%%%%%%%%%%%%%%%%%%%% 
% Results of Main Volume %
%%%%%%%%%%%%%%%%%%%%%%%%%% 
\begin{figure*}
    \hspace{1.0cm}{\large (a) Velocity power spectrum} 
    \hspace{3.0cm} {\large (b) Mode fraction }  \\
    \begin{minipage}[t]{0.47\textwidth}
        \centering{\includegraphics[scale=1.1]{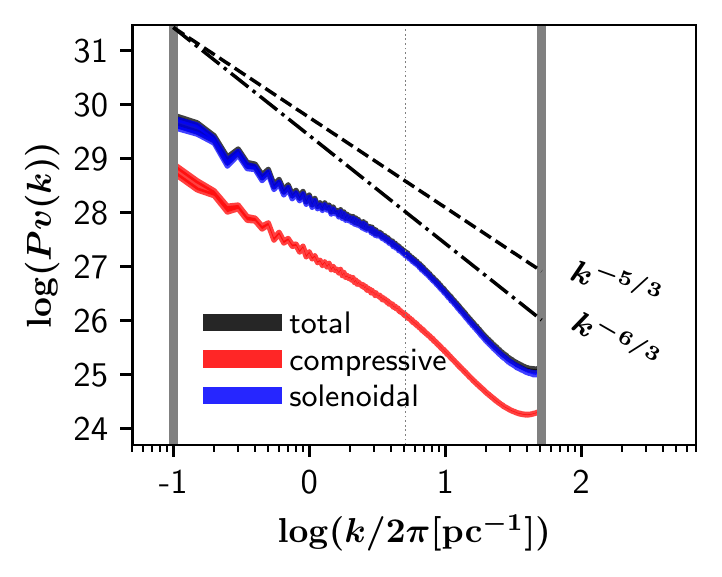}}
    \end{minipage}
    \hspace{0.5cm}
    \begin{minipage}[t]{0.47\textwidth}
        \centering{\includegraphics[scale=1.1]{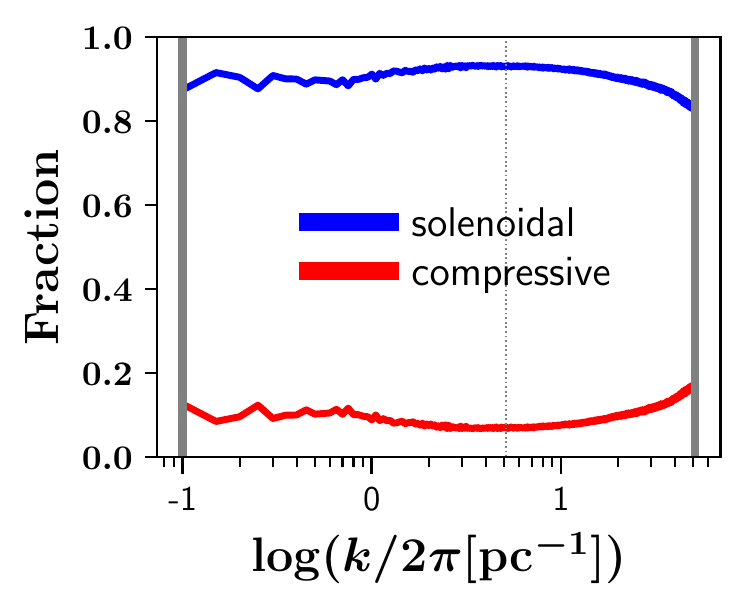}}
    \end{minipage}\\

    \caption{Panel (a): The one-dimensional averaged velocity power spectrum at 3 Myr over the entire volume of the shock-compressed layer. 
             The black line corresponds to the total power
             whereas the colored lines show the breakdown as the solenoidal modes (blue) and the compressive modes (red).
             The solenoidal mode amounts to $\gtrsim80$\% of the total power on all scales (shown in Panel (b)), 
             so that the ``total'' (black) is almost overlapped with ``solenoidal'' (blue) in this figure.
             The shade shows the Poisson noise based on the sample counts in each $k$-vector bin.
             The vertical gray lines show $k/2\pi = 0.1$ pc$^{-1}$ (left solid; the proxy of the lowest frequency of the shock-compressed layer 
             (see Appendix~\ref{sec:appA})),
             $k/2\pi = 5.12$ pc$^{-1}$ (middle thin dotted: the numerical diffusion likely impacts the power on $k/2\pi > 5.12$ pc$^{-1}$),
             and the Nyquist frequency $k/2\pi = 51.2$ pc$^{-1}$ (right solid). 
             We use the powers on $0.1$ pc$^{-1}$ $\leq k/2\pi \leq 5.12$ pc$^{-1}$
             in the analyses hereafter.
             We overplot two reference lines representing power-laws of $\propto k^{-5/3}$ and $k^{-6/3}$.
             The $k$-space is linearly binned where the highest frequency corresponds to the Nyquist frequency.
             Panel (b): Same as Panel (a) but the solenoidal/compressive mode fraction out of the total turbulent power.
             }
    \label{fig:PkF_all}
\end{figure*}

\section{Results I: The Entire Volume of the Shock-Compressed Layer}
\label{sec:allvol}
Hereafter, the spatial frequency $k/2\pi$ denotes the inverse of the wavelength of a corresponding wave,
such that  $k/2\pi =0.1$ pc$^{-1}$ corresponds to 10 pc.

\subsection{Solenoidal and Compressive Mode Fraction}
\label{subsec:scmf_allvol}
Panel (a) in Figure~\ref{fig:PkF_all} shows the one-dimensional averaged velocity power spectrum.
The overall spectrum follows the Kolmogorov spectrum $\propto k^{-5/3}$
where the solenoidal modes dominate than the compressive mode on all scales.
The spectrum of the solenoidal modes exhibit a break at $k/2\pi\sim 5.0$ pc$^{-1}$
so does the total spectrum.
This is attributed to the numerical diffusion, which we evaluate further in Appendix~\ref{sec:ndiff}.
Given that the typical cooling length of the UNM
peaks at $\sim 1$ pc 
(Panel (a) of Figure~\ref{fig:CNMhisto_ndiff} in Appendix~\ref{sec:ndiff})
and that 
the break of the compressive mode power is limited compared with 
that of the solenoidal mode power,
our spatial resolution is high enough to resolve the cooling length of the UNM
to follow the dynamical condensation through the thermal instability
(but also see Section~\ref{subsec:limit}).

Panel (b) in Figure~\ref{fig:PkF_all} shows the solenoidal/compressive mode fraction
out of the total turbulent power. This shows that the solenoidal modes
account for $80$--$90$ \% on most scales.

\begin{figure}
    \centering{\includegraphics[scale=0.95]{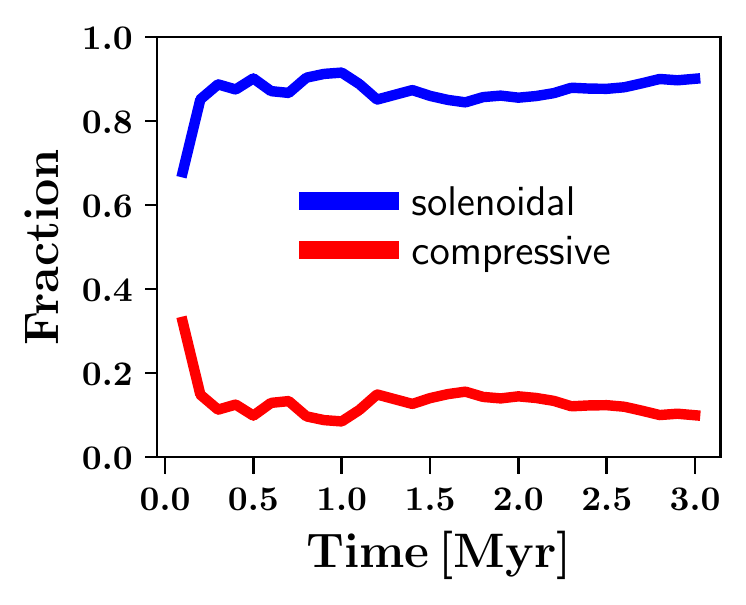}}
    \caption{The time evolution of the solenoidal/compressive mode fraction integrated over the entire volume of the shock-compressed layer.
             The blue (red) line shows the solenoidal (compressive) modes.
             We calculate the sum of the modes on $0.1$ pc$^{-1}$ $\leq k/2\pi \leq 5.12$ pc$^{-1}$
             to evaluate the fractions.
             At the finale timestep, 3 Myr, the fraction of the compressive and solenoidal modes are 
             $f_{\rm comp} = 0.10$ and $f_{\rm sol}=0.90$ respectively.
             The corresponding mode ratio $\chi$ becomes $0.11$ so that the forcing parameter is $b=0.32$.}
    \label{fig:tem}
\end{figure}

Figure~\ref{fig:tem} shows the time-evolution of
the total solenoidal/compressive mode fraction integrated from 
$k/2\pi = 0.1 $ to $5.12$ pc$^{-1}$ 
over the entire volume of the shock-compressed layer.
A converging flow configuration itself is a compressive motion 
and thus the compressive mode initially exists $\sim 40$ \%.
Along with the turbulent generation, the solenoidal modes increase
to reach $80$--$90$ \%.
After $0.5$ Myr, the fractions into the solenoidal and compressive modes
are quasi-steady with $80$--$90$ \%
and $10$--$20$ \%, respectively.
In this quasi-steady state, the mass fraction between WNM, UNM, and CNM 
is also steady, and the expansion speed of the shock-compressed layer 
is almost constant \citep{Kobayashi2020}.
At the final timestep, 3 Myr, 
the fraction of the compressive and solenoidal modes are 
$f_{\rm comp} = 0.10$ and $f_{\rm sol}=0.90$ respectively.
The corresponding mode ratio $\chi$ becomes $0.11$ so that the forcing parameter is 
$b=0.32$.

\begin{figure*}
    \hspace{1.0cm}{\large (a) The phase diagram} 
    \hspace{5.0cm} {\large (b) The density PDF}  \\
    \begin{minipage}[t]{0.48\textwidth}
        \centering{\includegraphics[scale=1.05]{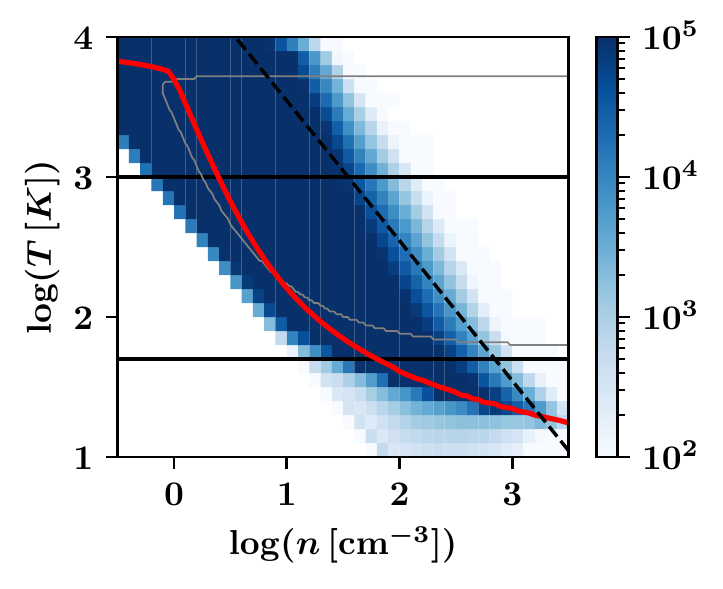}}
    \end{minipage}
    \hspace{0.5cm}
    \begin{minipage}[t]{0.48\textwidth}
        \centering{\includegraphics[scale=1.05]{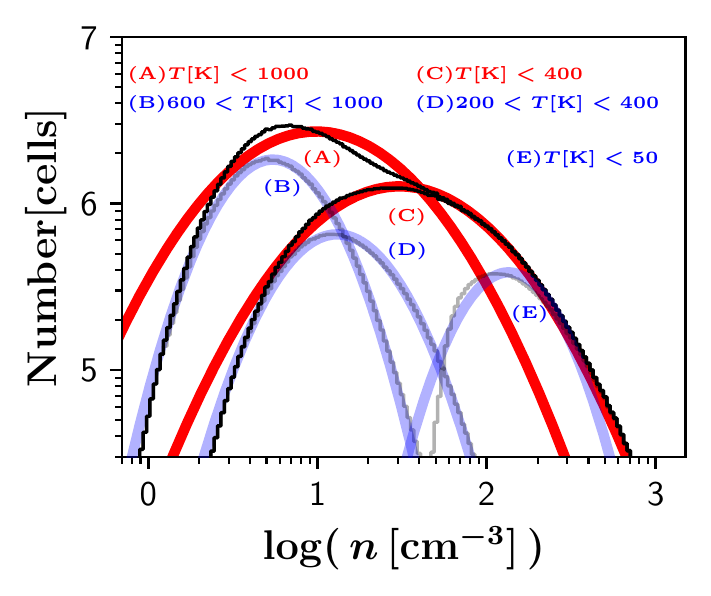}}
    \end{minipage}\\
    \caption{Panel (a): The cell histogram on the temperature-density diagram at 3 Myr.
             The color is coded as $\log$(Cells).
             The grey thin curve encloses the thermally unstable region
             with $(\partial \left(\mathcal{L}/T \right)/\partial T)_{P} < 0$
             \citep{Balbus1986,Balbus1995}.
             The red thick curve shows the thermal equilibrium state.
             The two horizontal black lines correspond to $T=1000$ K and $T=50$ K.
             The black dashed line shows the temperature corresponding to the ram pressure of the converging flow.
             Panel (b): The volume-weighted density PDF (\ie, cell histogram) and its fitted log-normal function 
             of the entire volume of the shock-compressed layer at 3 Myr. 
             We plot five histograms and corresponding log-normal functions 
             based on temperature thresholds; 
             $T[\mathrm{K}] \leq 1000$ (left red),
             $600< T[\mathrm{K}] \leq 1000$ (left blue),
             $T[\mathrm{K}] \leq 400$ (middle red),
             $200< T[\mathrm{K}] \leq 400$ (middle blue),
             $T[\mathrm{K}] \leq 50$ (right blue).
    }
    \label{fig:PDF00}
\end{figure*}

\subsection{Density PDF and its Width with Various Thermal States}
\label{subsec:dPDF_allvol}
Panel (a) of Figure~\ref{fig:PDF00} shows 
the thermal state within the shock-compressed layer at 3 Myr.
This shows that the shock-heated WNM at $n=1$-$10$ cm$^{-3}$ and $T\sim6000$ K
cools through the thermally unstable regime, 
and their densest parts reach the CNM state at $100$ -- $1000$ cm$^{-3}$ and $T\lesssim 50$ K.
The wide distribution on this phase diagram realizes due to the combination 
of the turbulent mixing and the thermal instability.
For example, if the converging flow is a completely one-dimensional head-on collision as $\sqrt{\langle \delta n_0^2 \rangle}/n_0 = 0$,
the medium cools with the constant pressure set by the ram pressure of the converging flow
(along with the black dashed line on Panel (a) of Figure~\ref{fig:PDF00}).

Panel (b) of Figure~\ref{fig:PDF00} shows 
the density PDF,
overplotted with the corresponding log-normal fittings\footnote{Note that, 
although the histograms are well fitted by log-normal functions,
the heating/cooling processes always induce deviation from the log-normal functions
because the temperature and density in the thermally-stable state have correlations.
For example, the density PDF of $T<50$ K has a strong transition 
at $50 - 100$ cm$^{-3}$, where the thermally-stable state has $T=50$ K
so that most of the gas with $T<50$ K is restricted to $n\gtrsim 100$ cm$^{-3}$
(see Panel (a) of Figure~\ref{fig:PDF00}).}.
Table~\ref{table:prop_00} summarizes the measured properties. 
The mean sound speed and the mean velocity dispersion are evaluated as
\begin{eqnarray}
    \langle \mycs (T_t) \rangle      &=& \frac{\sum_{T(\mathbf{x})\in T_t} \sqrt{\gamma \, P(\mathbf{x})/\rho(\mathbf{x}) }}{ \sum_{T(\mathbf{x})\in T_t}} \,, 
    \label{eq:sigma_Cs} \\
    \sqrt{ \langle \delta v^2 (T_t) \rangle} &=& \sqrt{ \frac{\sum_{T(\mathbf{x})\in T_t} \left( \mathbf{v}(\mathbf{x}) - \mathbf{\bar{v}}\right)^2 }{ \sum_{T(\mathbf{x})\in T_t} } }\,.
    \label{eq:sigma_vt}
\end{eqnarray}
Here $T(\mathbf{x})\in T_t$ indicates that 
we select cells whose temperature satisfies the temperature thresholds of $T_t$.
For this $T_t$ thresholds,
we select
$T\leq 50$ K, $T\leq 200$ K, $T\leq400$ K, $T\leq600$, and $T\leq1000$ K,
to cover the transition from the UNM down to the CNM.
We derive the mean Mach number as $\langle \mathcal{M}(T_t) \rangle = \sqrt{ \langle \delta v^2 (T_t) \rangle} / \langle \mycs(T_t) \rangle $.
$\myssim$ shows the width of the fitted log-normal distribution.
We also obtain $\mysiso$ by inserting the measured 
$b=0.32$ and $\langle \mathcal{M} \rangle$ into Equation~\ref{eq:ssiso} (Section~\ref{subsec:scmf_allvol}).
Note that measuring $b$ within individual temperature thresholds 
is difficult, especially cold components that are spatially disconnected (see Figure~\ref{fig:rhoT}).
Nevertheless, in Section~\ref{subsubsec:CNMstats}, we will utilize the velocity field in the real space to
measure the numerical-grid-scale turbulence mode fraction  
and show that the fraction of the compressive (solenoidal) mode powers are 23\% (77\%).
This is still solenoidal-mode dominated and is similar to the averaged $b=0.32$ that we obtain here on the entire shock-compressed layer.
In this section, therefore, 
we simply employ $b=0.32$ for all the temperature ranges 
when evaluating $\mysiso$.

\begin{table*}
    \caption{Measured properties from the entire volume of the shock-compressed layer at 3 Myr}
    \centering{
        \hspace{-2cm}\begin{tabular}{c||c|c|c|c|c}
            \\ \hline\hline                                                                  & $f_{\rm comp}$                    &  $f_{\rm sol}$                    &$\chi$                             & $b$                               &  \\
\cline{2-5}
                                                                                 & 0.10                         & 0.90                         & 0.11                         & 0.32                         &  \\                         \hline\hline\hline
                                                                                 & $T[\mathrm{K}] < 50$ \hspace{1cm} & $T[\mathrm{K}]<200$  \hspace{1cm} & $T[\mathrm{K}]<400$  \hspace{1cm} & $T[\mathrm{K}]<600$  \hspace{1cm} & $T[\mathrm{K}]<1000$     \\ \hline
$\langle C_{\rm s} \rangle \, [\mathrm{km\, s^{-1}}]$                            & 0.65            \hspace{1cm} & 0.98            \hspace{1cm} & 1.31            \hspace{1cm} & 1.64            \hspace{1cm} &  2.25               \\  
$\sqrt{ \langle \delta v^2 \rangle} \, [\mathrm{km\, s^{-1}}]$                   & 3.93            \hspace{1cm} & 4.23            \hspace{1cm} & 4.43            \hspace{1cm} & 4.62            \hspace{1cm} &  4.98               \\ 
$\mathcal{M} = \sqrt{ \langle \delta v^2 \rangle} / \langle C_{\rm s} \rangle $  & 6.02            \hspace{1cm} & 4.31            \hspace{1cm} & 3.37            \hspace{1cm} & 2.82            \hspace{1cm} &  2.21               \\ 
$\myssim$                                                                        & 0.61            \hspace{1cm} & 0.96            \hspace{1cm} & 1.12            \hspace{1cm} & 1.22            \hspace{1cm} &  1.13               \\ 
$\mysiso$                                                                        & 1.24            \hspace{1cm} & 1.02            \hspace{1cm} & 0.87            \hspace{1cm} & 0.76            \hspace{1cm} &  0.63               \\  \hline\hline

        \end{tabular}
    }\par
    \bigskip
            \textbf{Note.} 
            The top row: the compressive mode fraction $f_{\rm comp}$, the solenoidal mode fraction $f_{\rm sol}$,  the mode ratio $\chi$, and the forcing parameter $b$.
            The middle row: 
            The top line shows the temperature threshold. 
            The second line, $\langle C_{\rm s} \rangle$, corresponds to the mean sound speed
            and 
            the third line, $\langle \sigma_{\rm v} \rangle$, corresponds to the velocity dispersion (see Equations~\ref{eq:sigma_Cs} and~\ref{eq:sigma_vt}).
            The fourth line, $\mathcal{M}$, is the effective Mach number based on $\langle C_{\rm s} \rangle$ and $\langle \sigma_{\rm v} \rangle$.
            The fifth line, $\sigma_{\rm s,sim}$, is the width of the density PDF from our simulations.
            The sixth line, $\mysiso$, is the width of the density PDF expected from the isothermal turbulent theory,
            inserting $b=0.315$ and $\mathcal{M}$ (the fourth line) into Equation~\ref{eq:ssiso}.
    \label{table:prop_00}
\end{table*}
Table~\ref{table:prop_00} shows that 
1) 
the measured width of the density PDF $\myssim$ and the expected width $\mysiso$
are close to each other when we combine both the UNM and CNM components,
especially at $T<400$ K,
2) but the difference becomes larger towards the higher temperature range ($T<1000$ K)
and the lower temperature range ($T<50$ K)
up to a factor of $\sim 2$.
The difference is attributed to the evaluation of $b$ and $\mathcal{M}$ 
in $\mysiso$ (see Equation~\ref{eq:ssiso}).
In the following subsections,  
we will show that $\myssim$ and $\mysiso$ are close to each other 
when we take into account the thermal states properly
to select the consistent 
$\langle \mycs(T_t) \rangle$, $\sqrt{ \langle \delta v^2 (T_t) \rangle} $
and $\myssim$ responsible for each other,
thus the difference between $\myssim$ and $\mysiso$ 
is a natural consequence of the multi-phase nature of this system.

\subsubsection{Warmer Region with $T\sim 1000$ K}
\label{subsubsec:WarmT}
In the warmer regions,
$\myssim$ is wider than $\mysiso$.
$\sqrt{ \langle \delta v^2 (T_t) \rangle}$ and $\langle \mycs(T_t) \rangle$
are volume-weighted quantities so that 
they tend to depend more on diffuse (\ie, warmer) components
when we expand the temperature threshold to the warmer regime.
For example, $\mysiso$ at $T<1000$ K depends more on 
WNM and UNM at $T\sim 1000$ K that just starts to cool.
On the other hand, the measured $\myssim$ tries to take into account
the cooler UNM and CNM components as well ($T<400$ K), so that 
$\myssim$ at $T<1000$ K is wider than $\mysiso$\footnote{The total density PDF 
of the whole temperature range is bimodal between WNM and CNM ,
which can be seen in the thick black histogram in Panel (b) of Figure~\ref{fig:PDF00}
labeled ``(A)''.
The log-normal fitting is close to the peak of warmer components that have more volume than the cold components.
}.

In Panel (b) of Figure~\ref{fig:PDF00}, we additionally show the breakdown of density PDF,
such as $600<T\mathrm{[K]}<1000$ (labeled as (B)) and 
$200<T\mathrm{[K]}<400$ (labeled as (D)).
If we restrict ourselves to the distribution (B) alone,
we find that $\myssim(600<T\mathrm{[K]}<1000) =0.629$.
Meanwhile $\langle \mycs (600<T\mathrm{[K]}<1000)  \rangle =2.96\, \kms$ and $\sqrt{ \langle \delta v^2 (600<T\mathrm{[K]}<1000) \rangle} =4.10\, \kms$,
so that $\mathcal{M}(600<T\mathrm{[K]}<1000) =1.39$ and $\mysiso(600<T\mathrm{[K]}<1000) =0.51$.
This shows that $\myssim$ and $\mysiso$ is closer to each other at $600<T\mathrm{[K]}<1000$
rather than at the entire $T\mathrm{[K]}<1000$,
suggesting that the original values of $\sqrt{ \langle \delta v^2 \rangle}$, $\langle \mycs \rangle$,
$\mathcal{M}$ and corresponding $\mysiso$ at $T<1000$ K is weighted to warmer components.

\subsubsection{Cold Region with $T<50$K: CNM Clump Statistics}
\label{subsubsec:CNMstats}
\begin{figure*}
    {\large CNM clump statistics} \\
    \hspace{0.2cm} {\large (a) Size histogram } 
    \hspace{2.3cm} {\large (b) Mass histogram }
    \hspace{1.4cm} {\large (c) Size - velocity dispersion relation}  \\
    \hspace{-0.15cm}
    \begin{minipage}[t]{0.33\textwidth}
        \centering{\includegraphics[scale=0.84]{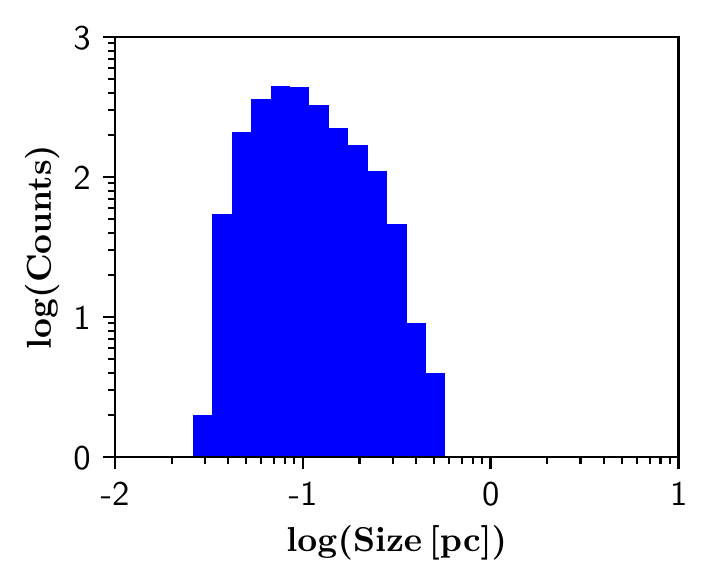}}
    \end{minipage}
    \hspace{-0.25cm}
    \begin{minipage}[t]{0.33\textwidth}
        \centering{\includegraphics[scale=0.84]{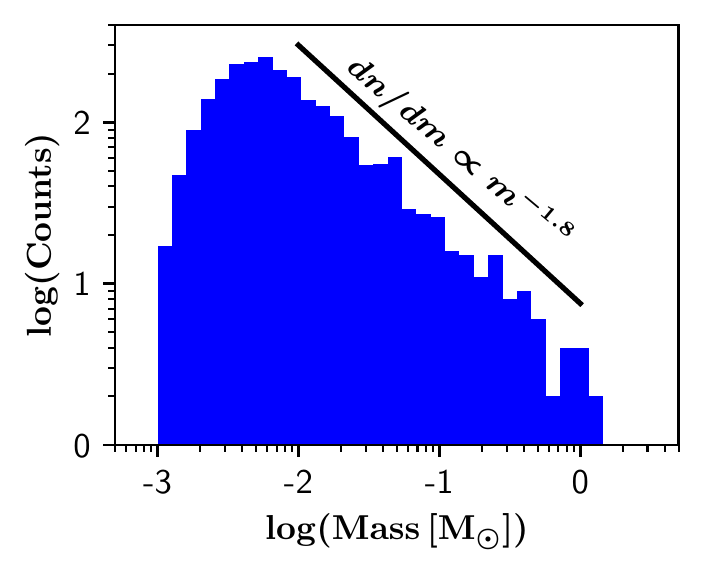}}
    \end{minipage}
    \hspace{-0.25cm}
    \begin{minipage}[t]{0.33\textwidth}
        \centering{\includegraphics[scale=0.84]{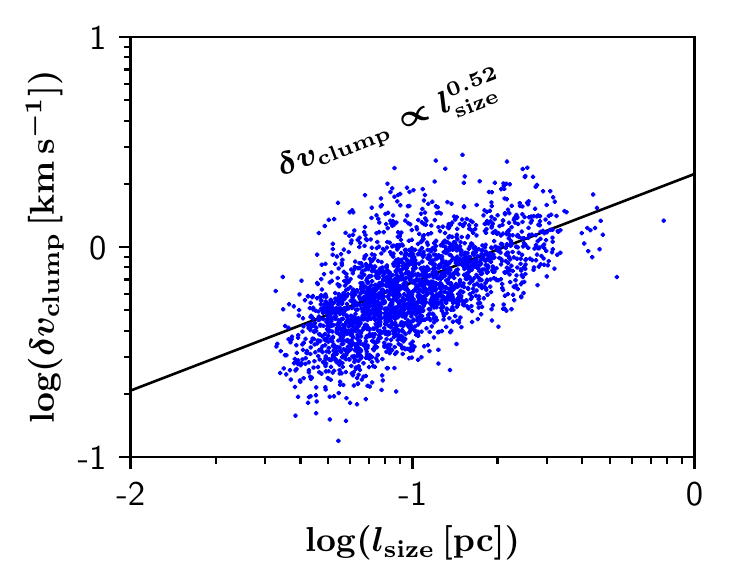}}
    \end{minipage}
    \caption{The statistics of 2403 CNM clumps:
             Panel (a) shows the size histogram, Panel (b) shows the mass histogram,
             and Panel (c) shows the size - internal velocity dispersion relation at 3 Myr.
             In Panel (c), each small circle represents individual CNM clumps
             and the black solid line is the least-square fitting result, 
             which is $\delta v_{\rm clump} \propto l_{\rm size}^{0.52}$.
             }
    \label{fig:CNMhisto_lb00}
\end{figure*}
Contrarily to the warmer regimes, 
$\mysiso$ is wider than $\myssim$
at colder regimes.
In $T<50$ K, for example,
$b\simeq0.1$ is required to have $\myssim = \mysiso$ with the given $\mathcal{M}=5.99$.
Such a small $b$ less than 0.3 is difficult to achieve in reality
even with a purely solenoidal turbulence \citep{Federrath2008}.
Figure~\ref{fig:PkF_all} already indicated that
Kolmogorov spectrum extends to smaller scales
with the almost constant $b \sim 0.3$,
even when we perform simulations with a higher resolution
to remove the numerical diffusion effect.
The generation of the compressive mode due to the thermal instability 
also takes place on small scales.
Therefore, our results suggest that the evaluation of $\mathcal{M}$ 
plays a more significant role to explain the difference 
between $\myssim$ and $\mysiso$.
As seen in Section~\ref{subsec:code},
the sub-pc-scale CNM clumps form through the thermal instability
and they are spatially disconnected to each other.
Our straightforward evaluation of $\sqrt{ \langle \delta v^2 (T_t) \rangle}$
Equation~\ref{eq:sigma_vt}, therefore reflects
the relative velocity between CNM clumps
more than the internal velocity dispersion within individual CNM clumps.

To further confirm the above discussions,
we identify CNM clumps and measure their physical properties.
We define a CNM clump as a connected volume 
with the temperature $T<50$ K,
which typically has its density $n\gtrsim50$ cm$^{-3}$ (see Panel (a) of Figure~\ref{fig:PDF00}).
In the following analyses,
we restrict ourselves to the clumps with $\geq 512$ cells
to reduce the effect of numerical noises,
and analyze 2403 such CNM clumps.

Panels (a) and (b) of Figure~\ref{fig:CNMhisto_lb00} show the size
and mass histograms at 3 Myr.
We use $\sqrt{I_{\rm max}/M}$ as the size of individual clumps,
where each clump has $I_{\rm max}$ as the maximum eigenvalue of its inertia matrix
and $M$ as its mass. The size and mass peaks at $\sim 0.08$ pc and $6 \times 10^{-3}\, \msun$.
The volume-weighted mean density of a clump\footnote{Hereafter, 
we use the subscript ``clump'' as the volume-weighted mean within each clump,
and $\langle \rangle_{\rm V}$ as the mean over all clumps weighted with each clump's volume.
For example of $\langle n_{\rm clump} \rangle_{\rm V}$,
we first measure the volume-weighted mean density of $i$-th clump,  $n_{{\rm clump},i}$,
then obtain the overall volume-weighted mean density as
$\langle n_{\rm clump} \rangle_{\rm V} = \sum_i n_{{\rm clump},i} V_i / \sum_i V_i$.
Here, $V_i$ is the volume of $i$-th clump and $i$ spans all CNM clumps.}
is $\langle n_{\rm clump} \rangle_{\rm V} \simeq 189$ cm$^{-3}$.
As the mass spectrum of the CNM clumps is $dn/dm \propto m^{-1.8}$, more massive clumps dominate the mass budget of CNM components.
This spectrum is often reported by other authors investigating converging flows \citep[\eg,][]{Inoue2012}
and consistent with a theoretically expected spectrum of $dn/dm \propto m^{-1.78}$ under the thermal instability growth
\citep[see \eg,][]{Hennebelle2007a}.

\begin{figure*}
    \hspace{0.2cm} {\large (a) $T[\mathrm{K}] < 50$ } 
    \hspace{3.0cm} {\large (b) $200 \leq T[\mathrm{K}] < 400$}
    \hspace{1.8cm} {\large (c) $600 \leq T[\mathrm{K}] < 1000$}  \\
    \begin{minipage}[t]{0.33\textwidth}
        \centering{\includegraphics[scale=0.80]{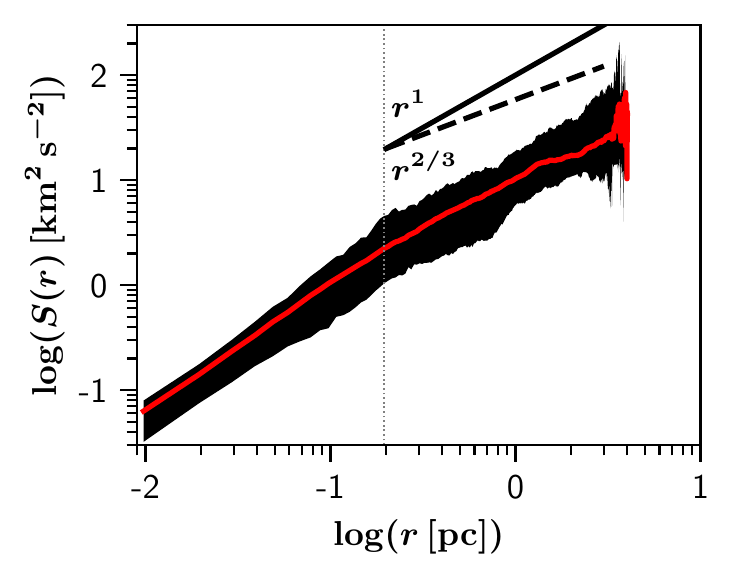}}
    \end{minipage}
    \hspace{-0.15cm}
    \begin{minipage}[t]{0.33\textwidth}
        \centering{\includegraphics[scale=0.80]{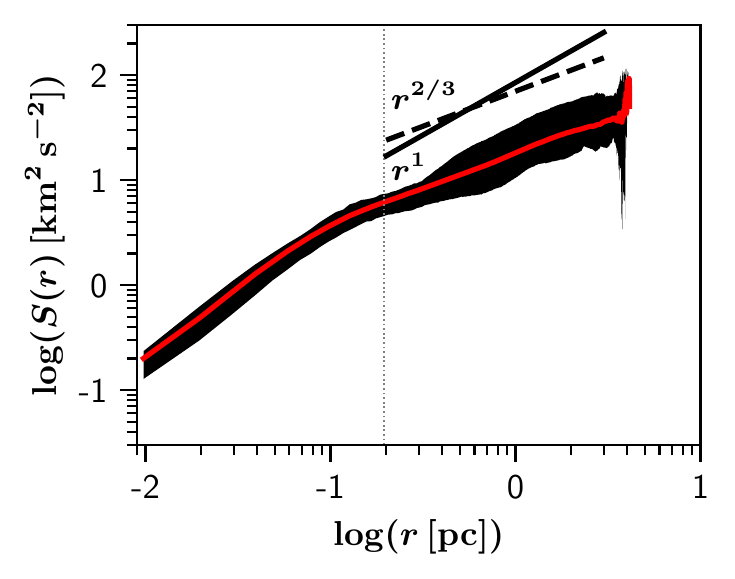}}
    \end{minipage}
    \hspace{-0.15cm}
    \begin{minipage}[t]{0.33\textwidth}
        \centering{\includegraphics[scale=0.80]{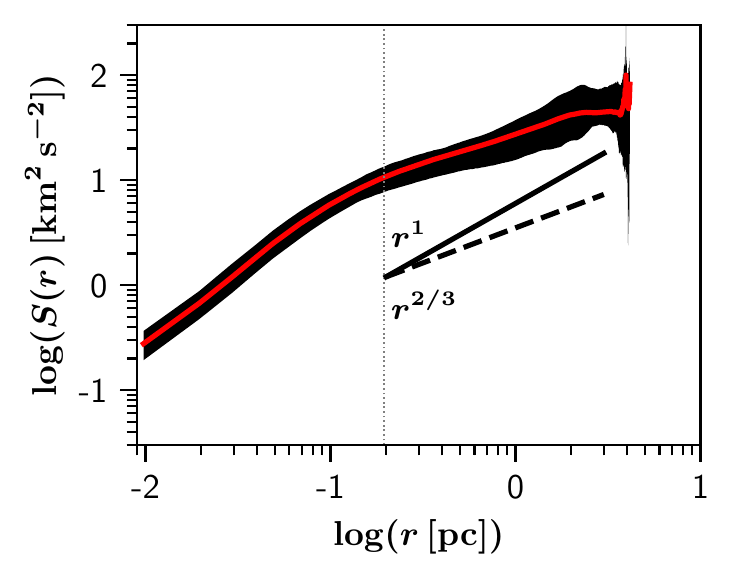}}
    \end{minipage}
    \caption{The second-order velocity function $S(r)$ averaged over the nine sub-volumes.
             Each panel corresponds to the temperature thresholds of
             (a) $T[\mathrm{K}] < 50$, (b) $200 \leq T[\mathrm{K}] <400$, (c) $600 \leq T[\mathrm{K}] < 1000$ at 3 Myr.
             The black shade shows the range of the measured $S(r)$ 
             whereas the lines show the power-law function
             of Kolmogorov (dashed: $S(r) \propto r^{2/3}$) and Larson (solid: $S(r) \propto r^1$).
             The red line shows the mean of nine sub-volumes.
             We discuss the powers on $r\gtrsim 0.2$ pc (shown as vertical gray dotted line),
             where the powers are free from the numerical diffusion (see Appendix~\ref{sec:ndiff}).
             }
    \label{fig:vsf}
\end{figure*}

Panel (c) of Figure~\ref{fig:CNMhisto_lb00} shows the relation between the size and internal velocity dispersion 
within each CNM clump at 3 Myr.
The mean internal velocity dispersion
and the mean sound speed are
$\sqrt{ \langle (\delta v^2)_{\rm clump}  \rangle_{\rm V}} = 1.05$ km s$^{-1}$
and $\langle C_{\rm s, clump} \rangle_{\rm V} = 0.43$ km s$^{-1}$,
so that the mean Mach number within CNM clumps is $\mathcal{M}_{\rm clump} \sim 2.44$.
This internal velocity dispersion is consistent with $\lesssim 1$ km s$^{-1}$ 
suggested from the second-order velocity structure function
on the 0.1 pc scale
(shown later in Section~\ref{subsec:ssf} as Figure~\ref{fig:vsf}).

In order to evaluate the mode parameter within CNM clumps,
$b_{\rm clump}$,
we measure the powers of the grid-scale turbulent modes
as $\langle ( \nabla_{\rm g} \cdot \mathbf{v} )^2_{\rm clump} \rangle_{\rm V}$ for the compressive mode
and $\langle ( \nabla_{\rm g} \times \mathbf{v} )^2_{\rm clump} \rangle_{\rm V}$ for the solenoidal mode \citep{Kida1990a,Kida1990b}.
Here $\mathbf{v}$ is the velocity field in the real space, $\nabla_{\rm g}$ is the spatial differential operator on
each numerical cell scale.
The fraction of the compressive (solenoidal) mode powers are 23 \% (77 \%),
and the corresponding mode parameter on the grid-scale within CNM clumps is $b_{\rm clump} \sim 0.48$.
$b_{\rm clump}$ is, therefore, indeed larger than 0.1 as we speculated in the first paragraph of this subsection.
$b_{\rm clump} \sim 0.48$ also suggests that the mode fraction at the Nyquist frequency alone
in the Fourier analysis (the grey vertical line in Panel (a) of Figure~\ref{fig:PkF_all})
already indicated a fraction close to $b_{\rm clump}$
even though such a Fourier power is dominated rather by the WNM and UNM than by the CNM.

Equation~\ref{eq:ssiso} with $\mathcal{M}_{\rm clump} \sim 2.44$
and $b_{\rm clump} \sim 0.48$
expects that the density PDF width of the CNM component alone 
is $\mysiso \simeq 0.61$,
which is consistent with $\myssim = 0.61$.

In conclusion of Sections~\ref{subsubsec:WarmT} and~\ref{subsubsec:CNMstats}, 
our results suggest that
1) when we focus on the volume with $T<1000$ K, 
turbulence is more powered by diffuse warm components,
such as the WNM and UNM that just start to cool\footnote{\cite{Inoue2012} also 
reported similar properties where the power of WNM and UNM turbulence is stronger than that of the CNM
in their molecular cloud formation simulations in the magnetized medium.},
thus the isothermal theory with the volume-weighted turbulence
explains the width of the density PDF of such warmer components
without the cold tails,
2) the coldest part of the density PDF ($T<50$ K)
is governed by the turbulence within individual CNM clumps
than by the strong turbulence on the entire cloud scale,
which determines the relative velocity between CNM clumps.
The turbulent $\mathcal{M}$ within individual CNM clumps
is more important to evaluate the density PDF of the CNM component alone,
and 3) Equation~\ref{eq:ssiso} from the isothermal theory 
is directly applicable when we compile all the volumes with $T<400$ K.

\begin{figure*}
    \hspace{1.0cm}{\large (a) Velocity power spectrum} 
    \hspace{3.0cm} {\large (b) Mode fraction }  \\

    \begin{minipage}[t]{0.47\textwidth}
        \centering{\includegraphics[scale=0.9]{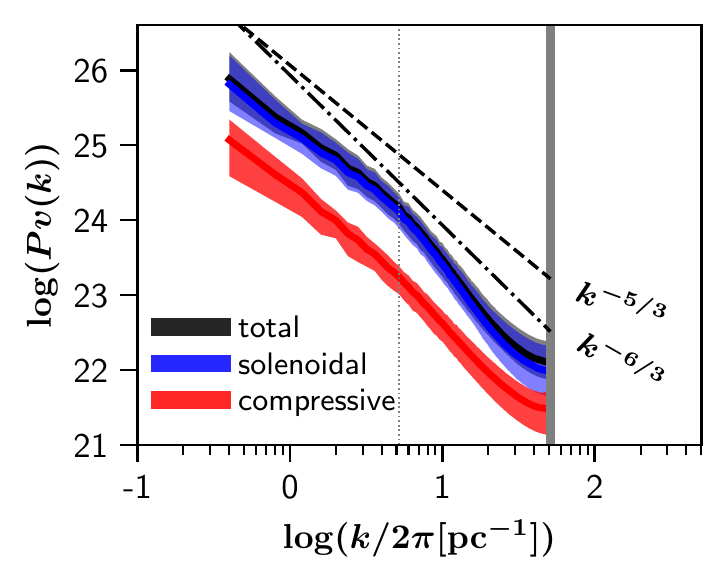}}
    \end{minipage}
    \hspace{0.5cm}
    \begin{minipage}[t]{0.47\textwidth}
        \centering{\includegraphics[scale=0.9]{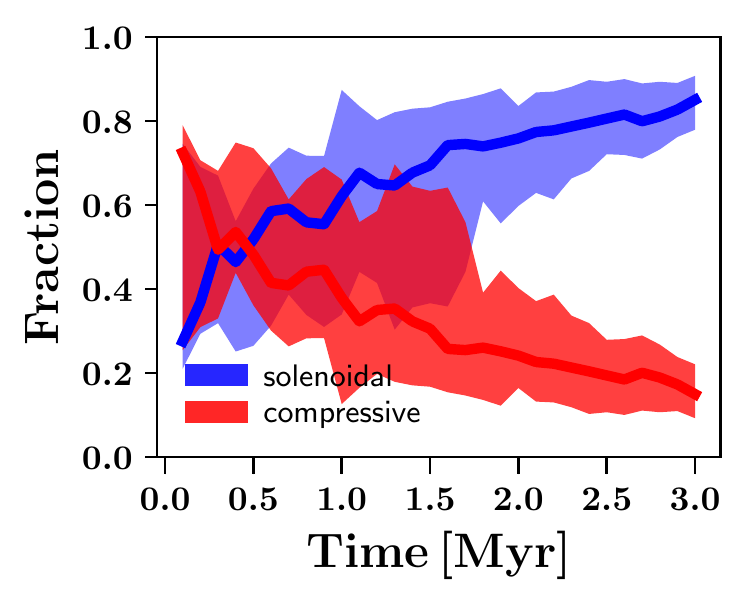}}
    \end{minipage}\\
    \caption{Panel (a): The velocity power spectrum compiled from the nine sub-volumes of the shock-compressed layer at 3 Myr. 
    The color represents the turbulence mode (black for the total, blue for the solenoidal modes, and red for the compressive modes).
    The shade corresponds to the minimum and maximum range of each mode among the sub-volumes.
    The thick colored lines within each shade shows the mean of the nine sub-volumes.
    The vertical gray lines show the Nyquist frequency $k/2\pi = 51.2$ pc$^{-1}$ (right solid),
    and $k/2\pi = 5.12$ pc$^{-1}$ (middle thin dotted: the numerical diffusion likely impacts the power on $k/2\pi > 5.12$ pc$^{-1}$).
    We overplot two reference lines representing power-laws of $\propto k^{-5/3}$ and $k^{-6/3}$.
    Panel (b): The time evolution of the solenoidal/compressive mode fraction compiled over nine sub-volumes.
    The blue (red) line shows the solenoidal (compressive) modes.
    We utilize the powers in on $0.39$ pc$^{-1}$ $\leq k/2\pi \leq 5.12$ pc$^{-1}$ for the analysis of the turbulent mode fraction.
    The shade corresponds to the minimum and maximum range among the sub volumes.
    }
    \label{fig:PkF_sub}
\end{figure*}
\begin{figure}
    \centering{\includegraphics[scale=1.0]{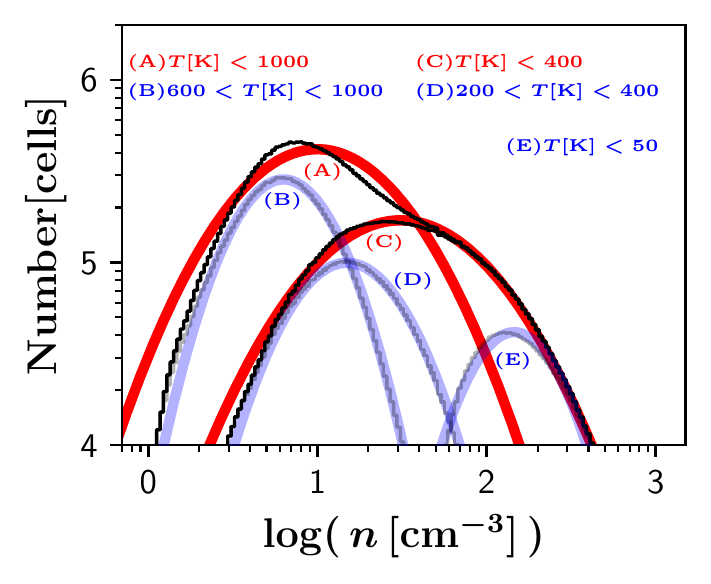}}
    \caption{The volume-weighted density PDF (histogram) and its fitted log-normal function 
             of the nine sub-volumes at 3 Myr. 
             }
    \label{fig:PDFb5}
\end{figure}

\begin{table*}
    \caption{Mean measured properties from the nine sub-volumes of the shock-compressed layer at 3 Myr}
    \centering{
        \hspace{-2cm}\begin{tabular}{c||c|c|c|c|c}
            \\ \hline\hline
                                                                                 & $f_{\rm comp}$                    &  $f_{\rm sol}$                      &$\chi$                                & $b$                                   &  \\
\cline{2-5}
                                                                                 & 0.15                             & 0.85                               & 0.18                                & 0.39                                 &                                                 \\ \hline\hline\hline

                                                                                 & $T[\mathrm{K}] \leq 50$          \hspace{1cm} & $T[\mathrm{K}]\leq200$          \hspace{1cm} & $T[\mathrm{K}]<400$              \hspace{1cm} & $T[\mathrm{K}]<600$               \hspace{1cm} & $T[\mathrm{K}]<1000$     \\ \hline
$\langle C_{\rm s} \rangle \, [\mathrm{km\, s^{-1}}]$                            & 0.65 $\frac{+0.02}{-0.01}$  \hspace{1cm} & 1.01 $\frac{+0.05}{-0.06}$ \hspace{1cm} & 1.37 $\frac{+0.10}{-0.10}$  \hspace{1cm} & 1.70 $\frac{+0.14}{-0.14}$   \hspace{1cm} & 2.32 $\frac{+0.24}{-0.20}$ \\  
$\sqrt{\langle \delta v^2 \rangle} \, [\mathrm{km\, s^{-1}}]$                    & 2.40 $\frac{+1.44}{-0.44}$  \hspace{1cm} & 2.75 $\frac{+1.44}{-0.64}$ \hspace{1cm} & 2.98 $\frac{+1.39}{-0.70}$  \hspace{1cm} & 3.19 $\frac{+1.34}{-0.76}$   \hspace{1cm} & 3.61 $\frac{+1.20}{-0.88}$ \\ 
$\mathcal{M} = \sqrt{ \langle \delta v^2 \rangle} / \langle C_{\rm s} \rangle $  & 3.66 $\frac{+2.23}{-0.76}$  \hspace{1cm} & 2.72 $\frac{+1.33}{-0.63}$ \hspace{1cm} & 2.17 $\frac{+0.87}{-0.50}$  \hspace{1cm} & 1.87 $\frac{+0.61}{-0.44}$   \hspace{1cm} & 1.56 $\frac{+0.33}{-0.39}$ \\ 
$\myssim$                                                                        & 0.54 $\frac{+0.10}{-0.14}$  \hspace{1cm} & 0.85 $\frac{+0.12}{-0.17}$ \hspace{1cm} & 1.04 $\frac{+0.12}{-0.14}$  \hspace{1cm} & 1.11 $\frac{+0.14}{-0.13}$   \hspace{1cm} & 1.02 $\frac{+0.24}{-0.17}$ \\ 
$\mysiso$                                                                        & 1.05 $\frac{+0.31}{-0.30}$  \hspace{1cm} & 0.86 $\frac{+0.25}{-0.28}$ \hspace{1cm} & 0.73 $\frac{+0.21}{-0.25}$  \hspace{1cm} & 0.65 $\frac{+0.17}{-0.23}$   \hspace{1cm} & 0.55 $\frac{+0.13}{-0.21}$ \\  \hline\hline

        \end{tabular}
    }\par
    \bigskip
            \textbf{Note.} Same as Table~\ref{table:prop_00} but for the mean from the nine sub-volumes within the shock-compressed layer. The $\pm$ shows the maximum/minimum range around the mean value.
    \label{table:prop_b5}
\end{table*}

\section{Results II: THE SUB-VOLUMES OF THE SHOCK-COMPRESSED LAYER}
\label{sec:subvol}
In this section, we analyze nine sub-volumes
to investigate the local variations.
The nine sub-volumes are selected so that 
their volume-centered position is 
$(L_x/2, L_y/2, L_z/2)$ for the most-centered one, and
$(L_x/2\pm2.56\, \mathrm{pc}, L_y/2\pm2.56\, \mathrm{pc},L_z/2\pm2.56\, \mathrm{pc})$
for the other eight ones.
We assume the periodicity in the Fourier analysis, 
and also expect that the boundary condition does not significantly affect 
the turbulent powers on the scales of $k/2\pi \leq 5.12$ pc$^{-1}$
(\cf, Section~\ref{subsec:appA2}).

\subsection{Turbulence Modes and Density PDF}
\label{subsec:scmf_subvol}
Panel (a) of Figure~\ref{fig:PkF_sub} shows the averaged velocity power spectrum from the nine sub-volumes,
with the minimum and maximum range measured among the sub-volumes.
Albeit the wavelength range affected by the numerical diffusion
have a signification fraction in these sub-volumes,
the overall power spectrum shows the Kolmogorov spectrum
and the solenoidal modes always dominate the turbulence power,
similar to the result of the entire volume of the shock-compressed layer
(Section~\ref{subsec:scmf_allvol}).
Panel (b) of Figure~\ref{fig:PkF_sub} shows the time-evolution of the solenoidal and compressive modes,
which shows that,
at later stages of $\gtrsim 1.7$ Myr,
the variation among the nine regions is $\lesssim 10$ \% and the solenoidal modes 
amount to $\sim80$ \% of the total turbulence power\footnote{Note that the 
significant variation at early phases is likely affected by 
the choice of analyzed volume, so that 
focusing on only after 1 Myr is fair when we discuss the variation 
between the sub-volumes; see Section~\ref{subsec:appA2}.}.
At the final timestep, 3 Myr, 
the fraction of the compressive and solenoidal modes are 
$f_{\rm comp} = 0.15$ and $f_{\rm sol}=0.85$ respectively.
$\chi$ becomes $0.18$ so that the forcing parameter is $b=0.39$.

Figure~\ref{fig:PDFb5} shows the sum of the density PDF from the nine sub-volumes,
overplotted with the corresponding log-normal fittings.
We summarize the measured properties on the Table~\ref{table:prop_b5}.
Similar to Table~\ref{table:prop_00},
$\myssim$ and $\mysiso$ are comparable when we combine both the UNM and CNM components,
especially at $T\lesssim 400$ K.

In conclusion, the power spectrum and the density PDF suggest that
the local variation of the mode fraction exists 
but limited at about a 10 percent level
so that the solenoidal modes are always dominant.

\subsection{Second-Order Velocity Structure Function}
\label{subsec:ssf}
To investigate the connection between 
the various thermal states in our simulations
and the observations of the ISM turbulence,
we measure the second-order velocity structure function
within individual temperature threholds:
\begin{equation}
    S(r) = \langle \left| \mathbf{v}(\mathbf{x}+\mathbf{r}) - \mathbf{v}(\mathbf{x}) \right|^2 \rangle \,,
    \label{eq:svsf}
\end{equation}
where $r = \left| \mathbf{r} \right|$.
This analysis does not require any stringent boundary condition compared with the power spectrum analysis,
and is capable of
directly analyzing individual thermal states even when the volume of the focused thermal states are clumpy and spatially disconnected to each other.
For example, the structure function of CNM component alone can cover
from the turbulence within CNM clumps to the relative motion between CNM clumps on 10 pc scales.
However, the internal velocity dispersion are affected by the numerical diffusion 
as we have discussed so far. Therefore, let us focus on only large scales $\gtrsim 0.2$ pc 
when we have quantitative discussions in the following.

Figure~\ref{fig:vsf} shows the measured structure function.
As a representative of different thermal states, we employ 
$T<50$ K, $200$ K $< T < 400$ K, and $600$ K $< T < 1000$ K
similar to the density PDF analyses.
The analyzed volume 
in each panel of Figure~\ref{fig:vsf}
is limited to the one within the corresponding temperature range,
so that Panel (a) represents the CNM 
and Panels (b) and (c) mostly represent the UNM\@.
We overplot the power-laws of Kolmogorov ($S(r) \propto r^{2/3}$)
and Larson ($S(r) \propto r^1$).
The figure indicates that 
the UNM components exhibit the spectrum close to the Kolomogorov
(\eg, 600 $<$ T[K] $<$ 1000),
while the CNM components exhibit the spectrum close to the Larson (\eg, $T < 50$ K).
Therefore, the early phase of molecular cloud formation
through converging flows exhibits the Larson's law in the CNM components,
which is consistent with observed supersonic turbulence \citep[\eg,][]{Larson1981,Heyer2004}
while the overall turbulence follows the Kolomogorov spectrum.
Note that, even in more evolved phases, molecular clumps formed by the converging flow 
exhibit the Larson type size-velocity dispersion relation \citep[\eg,][]{Inoue2012}.

These results suggest that,
if we observationally measure the density PDF and the second-order velocity structure function in multiple thermal states,
they can be utilized to test our understanding of molecular cloud formation 
controlled both by the turbulence and the thermal evolution shown in this article,
and provide estimations of the turbulent forcing parameter $b$
more consistent with the thermal evolution of molecular clouds.

\section{Discussions}
\label{sec:discussions}

\subsection{Comparison with Previous Studies and the Importance of Self-gravity}
\label{subsec:comps}
We calculate up to 3 Myr 
and the accumulated mass is limited to 
$2.6 \times 10^3 \msun$.
We ignore the self-gravity,
which is sub-dominant compared with the ram pressure of the converging flow
and the thermal/turbulent energy 
until $\sim 40$ Myr \citep[\cf,][]{Kobayashi2020}.
Our calculation therefore corresponds to early phases of the cloud formation and evolution.

Compared with our simulations, other previous studies focus on more later stages of the cloud evolution 
and show the impact of the self-gravity.
For example, \cite{Matsumoto2015} perform hydrodynamics simulations 
to calculate 3 Myr evolution of an isothermal molecular cloud, 
whose initial mean density (mass)
is already $10^2$ -- $10^3$ cm$^{-3}$ ($10^3$ -- $10^4\, \msun$)
and whose initial turbulence is with the solenoidal modes alone.
They report that
the compressive mode fraction remains $\lesssim 20$ percent
without the self-gravity, which is close to our value of 10 percent,
while the compressive mode generation due to 
the self-gravity is able to enhance the fraction more than 20 percent.
\cite{Kortgen2017} also perform cloud formation simulations
with heating and cooling
similar to our studies,
but focus more on massive clouds ($\sim 10^5\, \msun$)
on larger scale (256 pc).
They investigate 
a long-term evolution (until 30 Myr) 
by employing a refinement technique
with the spatial resolution from 62 pc
to 0.03 pc at highest,
and measure the turbulent driving parameter $b$.
Their results show that
the parameter $b$ evolves from
0.3 to 0.8 due to the self-gravitational collapse of the cloud.

Our current results of $b\sim 0.3$ is consistent with
the early stages of these previous studies.
It is, however, still important to 
investigate the evolution of self-gravitating clouds
by resolving the thermal instability 
coherently from the beginning to later stages $\sim 40$ Myr.
This is left for future studies.

\subsection{Limitations: Internal Velocity Dispersion of CNM Clumps}
\label{subsec:limit}
In Section~\ref{subsubsec:CNMstats},
we show that, to determine the density PDF of the CNM component, 
the internal velocity dispersion in individual CNM clumps is more responsible than
the relative velocity between the CNM clumps.
Higher resolution is still ideal 
to perform a more precise quantitative comparison
between $\myssim$ and $\mysiso$,
because the $\sigma_s$ dependence on the Mach number
comes in with $\mathcal{M}^2$ as shown in Equation~\ref{eq:PS},
and also because the spatially-smallest growing mode of the thermal instability
is limited by the size of numerical cells
and our current calculations are affected by the numerical diffusion
on the scales below CNM clumps $\lesssim 0.1$ pc.
Such calculations are left for future studies
but we expect that this will be important when compared with (future) observations
(see Section~\ref{subsec:ISMint}).

Nevertheless, our current simulations
already obtained the mean power of the internal turbulence good enough
to discuss its importance on the density PDF\@.
In Appendix~\ref{sec:ndiff}, we will investigate the resolution dependence 
to show that our current simulation with 0.1 pc resolution has the convergence.
For example, the internal velocity dispersion of CNM clumps is 
1.05, 1.02, and 1.44 km s$^{-1}$ when the spatial resolution is 0.01, 0.02, and 0.04 pc.

This convergence originates in the fact that the power of the internal turbulence is stronger on larger scales
(as suggested in Panel (a) of Figure~\ref{fig:vsf}).
This means that, in the evaluation of the velocity dispersion responsible for the density PDF,
it is important to resolve the eddies whose sizes are close to CNM clumps' sizes.
The CNM clumps' sizes ($\sim 0.1$ pc) are typically comparable to 
the spatial scale at which the numerical diffusion starts to play in our current simulation ($\sim 0.2$ pc).
The structure function at 0.1 pc in Panel (a) is still close the continuation of the  
$r^1$ scale-dependence from the larger scales, which is free from the numerical diffusion ($\gtrsim 0.2$ pc).
This suggests that the impact of the numerical diffusion is still limited for the turbulent eddy whose size is $\sim 0.1$ pc.

Therefore, although it is still ideal to perform simulations with a higher spatial resolution to investigate 
more precise scale-dependence of the velocity dispersion within CNM clumps,
we expect that such simulations do not significantly change our current conclusion that 
the internal velocity dispersion is important to control the density PDF 
rather than the inter-clump relative velocity. 
Our convergence study in Appendix~\ref{sec:ndiff}
suggests that resolving the CNM clump size with $\gtrsim$ 5 cells
is the typical criteria to have the convergence in CNM clump mean properties.

\subsection{Caveats: Other conditions and physical processes}
The flow velocity that we employ in the current simulations 
is 20 km s$^{-1}$, a typical velocity of the shock propagation 
at late stages of supernova remnant expansion.
However a faster velocity $\gtrsim 100$ km s$^{-1}$
also exists in galaxy-galaxy tidal interactions or mergers
\citep[\eg,][]{Fukui2017a,Maeda2021}.
Such fast flows generate
stronger shear motion in the shock-compressed layer 
and can change the turbulent mode ratio
and/or CNM mass fraction by preventing the thermal instability.
This is beyond our scope in this article,
but the dependence on flow velocity 
has to be investigated 
to understand more intense star-forming regions
\citep[\cf,][for the Mach number dependence of the velocity/density power spectrum and the density PDF]{Kim2005,Federrath2012}.

Magnetic fields also play important roles in 
cloud formation and evolution through the
magnetic pressure \citep[\eg,][]{Hennebelle2000,Inoue2008,Heitsch2009,Inoue2012,vazquezsemadeni2011,Iwasaki2019}
and the formation of filamentary structures that host protostar 
\citep{Andre2010,Inoue2013}.
The thermal instability is able to grow along the magnetic filed lines
to form multiphase structure.
For examples, \cite{Inoue2012} and \cite{Iwasaki2019}
show that, even in magnetized cases,
the converging flows are able to create highly turbulent 
two-phase molecular clouds like the results in this article,
especially when the angle between the converging flow direction and magnetic fields is close to parallel.
We therefore should 
investigate whether our understanding is valid in magnetohydrodynamics regime
by comparing the known relation of the density PDF width 
in magnetized isothermal gas \citep[\eg,][]{Federrath2012}.
This is beyond the scope at this moment, 
but we are planning to extend our simulations towards this direction.

It is also important to calculate chemical networks (either on-the-fly
or post-process) followed with a synthetic observation
to directly compare with molecular line observations.
The WNM and CNM volumes are, in principle, accessible 
by H{\sc I} and CO line observations,
whereas we expect that the intermediate temperature range of a few 100 K 
is an interesting frontier
traceable by the combination of the OH lines \citep[\eg,][]{Ebisawa2020}
and by the H{\sc I} line, with ALMA, SKA, ngVLA and so on.
Such calculations are left for future studies at this moment.

\subsection{Implications to Observational Estimations in Interstellar Solenoidal/Compressive Mode Ratio}
\label{subsec:ISMint}
Direct observations of the solenoidal/compressive mode ratio parameter $b$
in actual molecular clouds are difficult 
because observations are limited to the position-position-velocity space.
Nevertheless, there have been attempts to estimate the solenoidal/compressive mode ratio
by using Equation~\ref{eq:ssiso} (or its magnetized version)
based on the measured Mach number and the width of log-normal density PDF
\citep[\eg,][]{Brunt2010,Federrath2016,Sharda2021}.
For example, \cite{Federrath2016} find that 
typical molecular clouds have their $b$ greater than 0.4,
with a significant compressive mode contribution 
exceeding the natural mixing, 
whereas the G0.253+0.016 cloud at the Central Molecular Zone 
exhibits $b=0.22$, dominated by the solenoidal modes.
The authors suspect that the origin of this variation 
is in the difference of the driving mechanism;
the strong shear motion is invoked by its recent galactic-pericenter passage
by Sgr A* \citep{Longmore2013b},
which explains the low star formation efficiency at the Galactic Center
\citep{Longmore2013a}.

The exact value of $b$, however, 
needs careful analyses.
For example, \cite{Brunt2010} discuss how 
the sub-resolution structures 
below $0.1$ pc
impact the $b$ estimation
by assuming that the Larson's law 
continues to the infinitely small spatial scale 
\citep{Heyer2001},
and conclude that 
such structures induce the underestimation 
in $b$ by a factor of $\sim \sqrt{2}$
in the case of the Taurus molecular cloud,
and this has to be tested by observations with a higher angular resolution.
Such sub-resolution effect is indeed important
because the thermal instability grow
on wide spatial ranges even below $0.1$ pc \citep{Koyama2000}.
Furthermore, 
even with a high angular resolution, 
observational estimations 
of the Mach number using molecular lines (\eg, CO, HNCO, \etc,)
have a chance to probe the relative velocity
between CNM clumps along the line of sight.
This relative velocity is powered more by the strong turbulence 
of the WNM and UNM
and does not determine the density PDF at the CNM temperature/density.
Therefore, to address the variation of the mode ratio parameter $b$ across various star-forming environments, 
we have to take 
the multiphase nature of the ISM into account,
by selecting the turbulent motion within the dense structures.
Higher spectral resolution is important 
as well as higher spatial resolution
(see also \citealt{Federrath2013b} for the importance of the Mach number
estimation as an origin of the variation in SFR at a given gas column density).

\section{Summary}
\label{sec:concl}
Towards the understanding of the turbulence properties realized in
the multiphase interstellar medium (ISM),
we perform a series of hydrodynamics simulations of converging warm neutral medium (WNM) flows
on a 10 pc scale with heating and cooling by which the thermally unstable neutral medium (UNM)
and the clumpy cold neutral medium (CNM) form.
We list our main findings as follows.

\begin{enumerate}
    \item The overall velocity power spectrum follows the Kolmogorov law
              where the solenoidal (compressive) modes account for $>80$ \% ($<20$ \%) 
              of the turbulence power on all scales.
    \item When we consider both the UNM and CNM components, the density probability distribution function (PDF) 
              has its width, $\myss$, close to the one expected from the framework of the isothermal turbulence theory
              as $\mysiso = \ln(1+b^2\mathcal{M}^2)$
              (where $\mathcal{M}$ and $b$ are the turbulent Mach number and the turbulence parameter representing the turbulence mode ratio, 
              respectively).
              Meanwhile, the width of the CNM component alone is narrower than the width that the isothermal theory predicts,
              if the isothermal theory employs the inter-clump relative velocity between CNM clumps
              as the main driver of turbulence responsible to control the CNM density PDF.
    \item These results suggest that
              the coldest part of the density PDF (\ie, clumpy CNM structures)
              is governed by the thermal instability and the resultant weak turbulence of $\lesssim 1$ km s$^{-1}$ within individual CNM clumps
              rather than by the strong turbulence on the entire cloud scale with $2$ -- $5$ km s$^{-1}$, which determines the relative velocity between CNM clumps.
              Observational interpretations of the CNM density PDF requires careful analyses to evaluate the turbulence
              within each CNM clump but not the relative velocity between CNM clumps; otherwise the Mach number of the turbulence
              can be overestimated and the corresponding forcing parameter $b$ is underestimated accordingly.
              $\mysiso$ is indeed close to $\myss$,
              when the isothermal theory uses \sout{we use} the mean Mach number of the internal turbulence within CNM clumps to derive $\mysiso$.
    \item The second-order velocity structure function suggests that 
              the overall volume follows the Kolmogorov spectrum, 
              as already shown by Figure~\ref{fig:PkF_all},
              and that the limited volume of the CNM component alone 
              follows more to the Larson's spectrum.
              If we observationally measure the density PDF and the second-order velocity structure function
              in multiple thermal states, they provide estimations of the turbulent forcing parameter $b$
              more consistent with the thermal evolution of molecular clouds.
    \item Our resolution study suggests that 
              resolving the CNM clump size with $\gtrsim$ 5 cells is the typical criteria
              to have the convergence in the mean properties of CNM clumps.
\end{enumerate}

These results have to be further investigated by 
including magnetic fields.
We also hope that 
upcoming observations (\eg, ALMA, SKA, ngVLA)
measure the UNM/CNM (column) density PDF
and the velocity structure function
with multiple lines (H{\sc i}, OH, CO)
to investigate our results.

\section*{ACKNOWLEDGMENTS}
Numerical computations were carried out on Cray XC30 and XC50 
at Center for Computational Astrophysics, National Astronomical Observatory of Japan.
MINK (15J04974, 18J00508, 20H04739), TI (18H05436, 20H01944),
KT (16H05998, 16K13786, 17KK0091, 21H04487),
and KI (19K03929, 19H01938), 
are supported by Grants-in-Aid from the Ministry of Education, Culture,
Sports, Science, and Technology of Japan. 
MINK appreciate Atsushi J. Nishizawa and Chiaki Hikage for helping our Fourier analysis.
MINK is grateful to 
Tomoaki Matsumoto, Kazuyuki Omukai, Hajime Susa, Sho Higashi, Gen Chiaki, Hajime Fukushima, Shu-ichiro Inutsuka,
Shinsuke Takasao, Tetsuo Hasegawa, and Kengo Tachihara for fruitful comments.
We are grateful to Hiroki Nakatsugawa who contributed the early phase of this study
through his master thesis work.

\appendix
\section{Numerical Diffusion and Ideal Spatial Resolution}
\label{sec:ndiff}
\begin{figure*}
    \hspace{1.0cm}{\large (a) With heating and cooling} 
    \hspace{4.0cm} {\large (b) Adiabatic with $\gamma=1.05$}  \\
    \begin{minipage}[t]{0.47\textwidth}
        \centering{\includegraphics[scale=0.9]{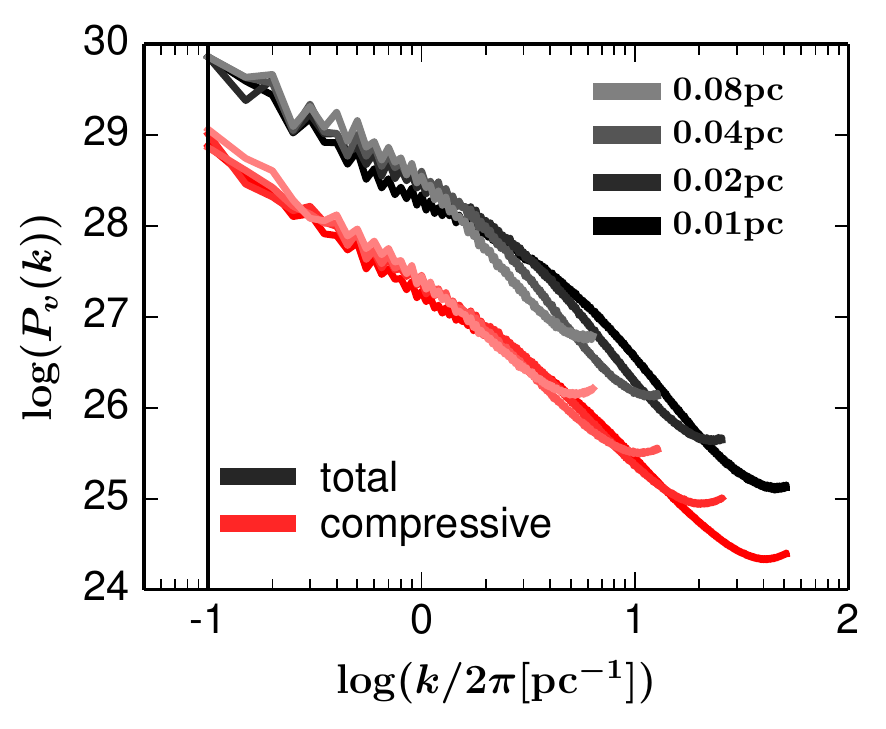}}
    \end{minipage}
    \hspace{0.5cm}
    \begin{minipage}[t]{0.47\textwidth}
        \centering{\includegraphics[scale=0.9]{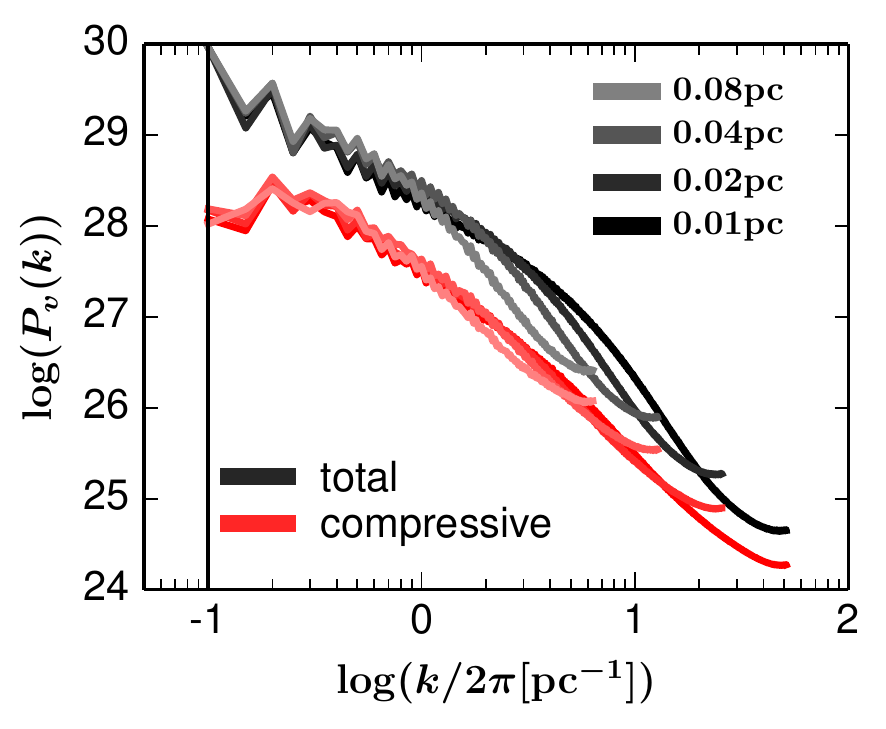}}
    \end{minipage}\\

    \caption{The velocity power spectrum for our fiducial calculation with heating and cooling 
             (Panel (a)), and that for our reference calculation with adiabatic converging flow
             of $\gamma=1.05$ (Panel (b)).
             The black lines represent the total power and the red lines represent the power of the compressive mode.
             The four lines correspond to the spatial resolution of 0.08 pc, 0.04 pc, 0.02 pc, and 0.01 pc.
             For the visualization purpose, we do not plot the solenoidal modes here, whose power
             is almost same as the total power.
             Panel (a) is at 3 Myr, whereas Panel (b) is at 1.5 Myr;
             the shock-compressed layer hits the $x$-boundaries at $\simeq 2$ Myr
             in the adiabatic case,
             so that for Panel (b) we choose the snapshot early enough to avoid the effect of the boundary.
             }
    \label{fig:Pk_comp}
\end{figure*}
As shown in Section~\ref{subsec:scmf_allvol}, the turbulence power spectrum exhibits a break
at an intermediate spatial scale of $k/2\pi \sim 5.12$ pc$^{-1}$.
Such a break widely appears in various previous simulations as well,
which is likely due to numerical diffusion;
\eg, \citealt[]{Federrath2013,Padoan2016};
also \citealt[]{Hennebelle2007a} (high-resolution albeit two-dimensional).
To investigate the possible responsibility of the numerical diffusion,
we perform additional simulations with multiple resolutions
and compare the velocity power spectrum, which is shown in Figure~\ref{fig:Pk_comp}.
Panel (a) shows the spectrum from fiducial runs with multiple spatial resolutions
and Panel (b) shows the spectrum from reference runs 
where we 
calculate the adiabatic converging flows with the specific heat ratio of $1.05$.
The choice of $1.05$ here aims to investigate
the evolution of the almost isothermal ISM, free from the heating and cooling effect.
These panels show that the break scale shifts at about one-tenth of the Nyquist frequency
as we change the spatial resolution,
suggesting that this decrease of the turbulence power at high frequencies 
is due to the numerical diffusion.
The relative increase of the compressive modes below this break scale
is due to the numerical diffusion accordingly.
We therefore employ powers above this break scale to evaluate 
the turbulence mode ratio $\chi$ and the forcing parameter $b$.
For example, when we analyze the entire volume of the shock-compressed layer in Section~\ref{sec:allvol},
we integrate the turbulence power from $k/2\pi = 0.1 $ to $5.12$ pc$^{-1}$.
Note that, in general, resolving vortex motions likely requires more numerical cells
than resolving contracting/expanding motions
so that the solenoidal modes tend to loose its power
than the compressive mode. 

\begin{figure*}
    \hspace{0.1cm} {\large (a) Effective cooling length} 
    \hspace{0.9cm} {\large (b) 0.02 pc resolution}
    \hspace{2.0cm} {\large (c) 0.04 pc resolution}  \\
    \begin{minipage}[t]{0.32\textwidth}
        \centering{\includegraphics[scale=0.80]{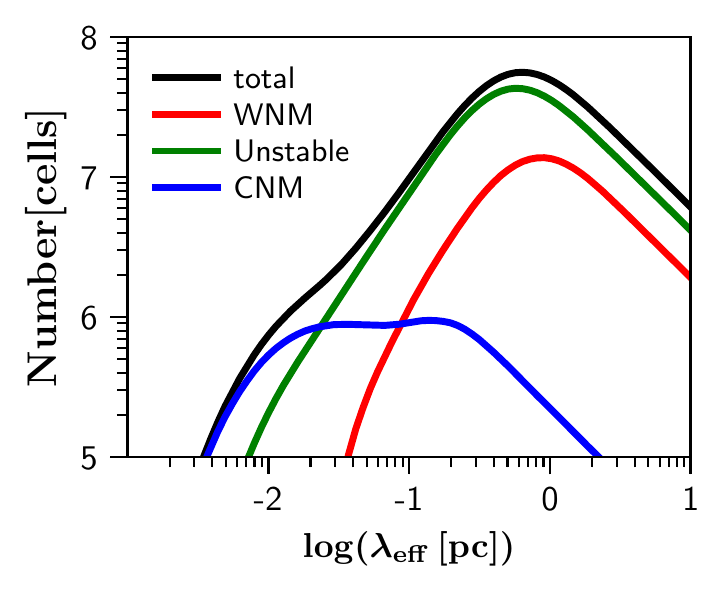}}
    \end{minipage}
    \hspace{-0.1cm}
    \begin{minipage}[t]{0.32\textwidth}
        \centering{\includegraphics[scale=0.80]{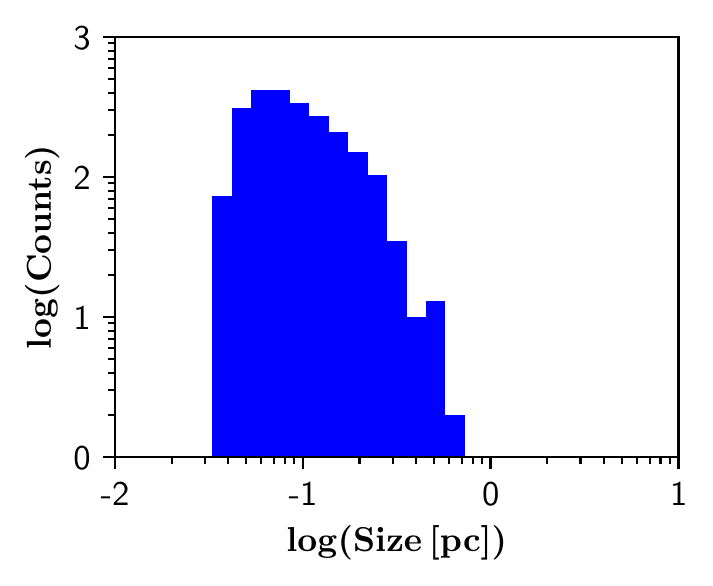}}
    \end{minipage}
    \hspace{-0.1cm}
    \begin{minipage}[t]{0.32\textwidth}
        \centering{\includegraphics[scale=0.80]{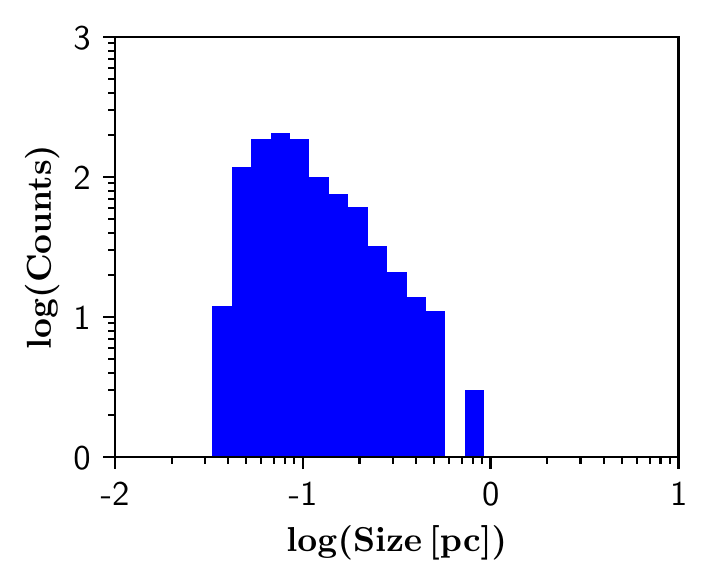}}
    \end{minipage}
    \caption{Panel (a): The histogram of the effective cooling length 
             $\lambda_{\rm cool,eff} =e \mycs/(\rho \mathcal{L} + P \nabla_i \mathbf{v}_i)$ at 3 Myr.
             The black shows the total histogram whereas the red/green/blue lines represent the histogram of the WNM/UNM/CNM\@. 
             Panels (b) and (c): The CNM clump size histogram with 0.02 pc spatial resolution (b) and with 0.04 pc spatial resolution (c).
             }
    \label{fig:CNMhisto_ndiff}
\end{figure*}
In our previous study \citep{Kobayashi2020}, 
we show that our current spatial resolution of 0.01 pc is 
already high enough compared with the typical cooling length
of the UNM $\sim 1$ pc and it is capable of following the dynamical condensation motion 
from the UNM to the CNM due to the thermal instability. 
Indeed \cite{Kobayashi2020} show that the CNM mass fraction is converged with the spatial resolution of 0.02 pc or higher.
Panel (a) of Figure~\ref{fig:CNMhisto_ndiff} shows the histogram of the effective cooling length $\lambda_{\rm cool,eff}$
in our current simulation, where we calculate $\lambda_{\rm cool,eff} = e \mycs/(\rho \mathcal{L} + P \nabla_i \mathbf{v}_i)$
in each numerical cell.
The transition from the WNM to CNM appears at $\sim 0.1$ pc,
which is consistent with the peak of CNM clumps' sizes in Panel (a) of Figure~\ref{fig:CNMhisto_lb00}.
Panel (a) of Figure~\ref{fig:CNMhisto_ndiff} suggests that 
the spatial resolution of 0.01 pc is a scale where the CNM dominates the cooling length.
This panel also suggests that the largest scale of the numerical diffusion ($\sim 0.2$ pc) resolves the typical cooling length of the UNM ($\sim 1$ pc),
and is close to the transition from the WNM to CNM\@.
Therefore, our current simulations indeed follow the dynamical condensation motion from the WNM/UNM to the CNM.

Our current 0.01 pc resolution, however, limits the highest frequency mode of the thermal instability growth.
CNM clumps are typically resolved with 10 cells per dimension (\ie, $0.1$ pc $/ 0.01$ pc), 
thus 
it is difficult to follow the detailed evolution of the turbulent structure within 
individual clumps.
In a self-gravitating system where the gravitational collapse 
enhances the turbulence, 
resolving the Jeans length by more than 32 cells 
is suggested as the criteria to have the convergence in the turbulent power
of the solenoidal modes \citep[\eg,][]{Federrath2011,Higashi2021}.
Our simulations presented in this article do not include self-gravity
but include heating/cooling processes, which invoke the 
dynamical condensation motion by the thermal instability 
and resultant turbulence on smallest scales.
The gas temperature decreases during this condensation motion
so that the 32-cell Jeans criterion of isothermal collapsing gas
is not directly applicable, 
but this criterion suggests that
we ideally should have $\sim 3 \times 10^{-3}$ pc or higher spatial resolution
to fully resolve the turbulent motions within individual CNM clumps of $\sim 0.1$ pc sizes.
In our current simulations, the fastest growing mode of the thermally instability 
should exist at the scale $\sim 7 \times 10^{-3}$ pc,
based on the typical sound speed (0.65 km s$^{-1}$) and the cooling time (0.011 Myr) of CNM clumps with 40 K and 189 cm$^{-3}$.
The spatial resolution $\lesssim 3 \times 10^{-3}$ pc is, therefore, desired to fully follow the formation of tiny CNM clumps
or tiny structure within CNM clumps on the scale of $\sim 7 \times 10^{-3}$ pc.
Such a high resolution simulation is difficult to perform in three dimension 
unless we ignore the cloud-scale evolution $\sim 10$ pc (\ie, WNM and UNM components)
and focus on local structures within molecular clouds,
which is left for future studies.

Nevertheless, the mass spectrum in Panel (b) of Figure~\ref{fig:CNMhisto_lb00} shows that the CNM mass budget is dominated by massive/large clumps so is the density PDF\@.
This is consistent with theoretically expected mass spectrum
under the thermal instability \citep{Hennebelle2007a}
and we expect that this spectrum is maintained even in simulations with much higher spatial resolutions.
Therefore, our current simulations already measure 
the turbulence most responsible for the density PDF,
which are eddies on the CNM clump scale (see Section~\ref{subsec:limit}).
In addition, we perform the CNM clump identification 
on the reference runs with the spatial resolution with 0.02 pc and 0.04 pc.
The minimum number of member cells is 64 cells (8 cells) for 0.02 pc (0.04 pc) resolutions,
scaled with the resolution.
Table~\ref{table:comp_resol} shows the measured CNM clump properties.
These results show that physical properties of the CNM clumps
as well as the grid-scale ratio of the turbulence mode 
start to have convergence with $\lesssim 0.02$ pc resolution.
This suggests that resolving the CNM clump size with $\gtrsim$ 5 cells (\ie, 0.1 pc / 0.02 pc)
is a typical criteria to have the convergence in CNM clump mean properties.
\begin{table}
    \caption{Resolution dependence of the CNM clump properties}
    \centering{
        \hspace{-2cm}\begin{tabular}{c||c|c|c}
            \\ \hline\hline
Resolution [pc]                                                                                                                   & 0.01  &  0.02  &  0.04   \\
\cline{1-4}
Number of CNM clumps                                                                                                              & 2403  &  2360  &  1028   \\ \hline\hline\hline
$\langle n_{\rm clump} \rangle_{\rm V}   \,  [\mathrm{cm^{-3}}] $                                                                 & 189   & 183    & 170     \\
$\langle C_{\rm s,clump} \rangle_{\rm V} \, [\mathrm{km\, s^{-1}}]$                                                               & 0.65  & 0.65   & 0.67    \\  
$\sqrt{ \langle (\delta v^2)_{\rm clump} \rangle_{\rm V} } \, [\mathrm{km\, s^{-1}}]$                                             & 1.05  & 1.02   & 1.44    \\ 
$\mathcal{M_{\rm clump}} = \sqrt{ \langle (\delta v^2)_{\rm clump} \rangle_{\rm V} } / \langle C_{\rm s,clump} \rangle_{\rm V} $  & 1.61  & 1.57   & 2.15    \\ 
$f_{\langle (\nabla \mathbf{v})^2        \rangle}$ [\%]                                                                           & 22.5  & 18.9   & 15.3    \\
$f_{\langle (\nabla \times \mathbf{v})^2 \rangle}$ [\%]                                                                           & 77.5  & 81.1   & 84.7    \\ \hline\hline

        \end{tabular}
    }\\
    \bigskip
            \textbf{Note.} 
            The top row: the spatial resolution and number of identified CNM clumps.
            The middle row: 
            The first line, $\langle n_{\rm clump} \rangle_{\rm V}$  shows the mean number density of a CNM clump.
            The second line, $\langle C_{\rm s,clump} \rangle_{\rm V}$  corresponds to the mean sound speed of a CNM clump.
            The third line, $\sqrt{ \langle (\delta v^2)_{\rm clump} \rangle_{\rm V} } $, corresponds to the mean internal velocity dispersion of a CNM clump.
            The fourth line, $\mathcal{M_{\rm clump}}$, is the mean effective Mach number within CNM clumps. %based on $\langle C_{\rm s} \rangle$ and $\langle \sigma_{\rm v} \rangle$.
            The fifth and sixth lines, $f_{\langle (\nabla \mathbf{v})^2        \rangle}$ ($f_{\langle (\nabla \times \mathbf{v})^2 \rangle}$) , 
            show the grid-scale fraction of compressive (solenodail) mode out of the total turbulence power, based on the measurements of
            $\langle ( \nabla_{\rm g} \cdot \mathbf{v} )^2_{\rm clump} \rangle_{\rm V}$ and $\langle ( \nabla_{\rm g} \times \mathbf{v} )^2_{\rm clump} \rangle_{\rm V}$. 

    \label{table:comp_resol}
\end{table}

\section{Assumptions in The Fourier Transformation and Technical Procedures to Derive the Power Spectrum}
\label{sec:appA}
In the evaluation of the mode ratio $\chi$ and the forcing parameter $b$,
we choose one-tenth of the Nyquist frequency as the highest frequency (\ie, $k/2\pi = 5.12$ pc$^{-1}$),
below which is the turbulent inertial range.
Meanwhile, the choice of the lowest frequency and the estimation of corresponding power on that scale
is not trivial.
In this section, we summarize the assumptions and the technical procedures involved in these analyses.

\subsection{The Frequency Range: Entire Shock-Compressed Layer}
\label{subsec:appA1}
As already shown in Figure~\ref{fig:rhoT}, the shock-compressed layer is significantly deformed,
which evolves in time, and thus performing the Fourier transformation on
such an arbitrary volume is not trivial.
Ideally we should apply some methods to remove the effect of such geometry. 
There are already several methods proposed in astrophysical context,
such as the Cosmic Microwave Background analysis
whose mask region onto the galactic foreground has the complicated shapes
\citep[\eg,][]{Planck2013_23},
or such as large galactic surveys 
where the observed area does not have any periodicity 
(\eg, Pseudo-spectrum method; \citealt{Brown2005,Hikage2011,Hikage2019}).
However, these high accuracy is beyond our scope at this moment. %in this article

We instead opt to fill the velocity fields outside the shock-compressed layer
uniformly by the mean bulk velocity of the shock-compressed layer,
so that the power of low frequencies are reduced,
which are otherwise dominated by the converging flow itself.
The converging flow configuration naturally results in
the mean bulk velocity of the shock-compressed layer as 0 
(in other words, the shock-compressed layer is by definition in the rest-frame of the contact discontinuity).
We therefore use $\mathbf{v}=0$ as this temporally filling velocity outside the shock-compressed layer
and perform the Fourier transform over the entire simulation domain.
Before the actual applications, we perform simple demonstrations.
We first generate turbulence of solenoidal-mode-only (compressive-mode-only) 
in 10 pc $\times$ 10 pc $\times$ 10 pc domain
by superposing 32 sinusoidal modes with the Kolmogorov power spectrum.
We locate this velocity field at the center of 20 pc $\times$ 10 pc $\times$ 10 pc domain
with the velocity field outside the central (10 pc)$^3$ be 0.
We perform the Fourier transform over the entire volume of 20 pc $\times$ 10 pc $\times$ 10 pc domain
and found that the power on the $k/2\pi = 1/20$ pc$^{-1}$ scale is successfully suppressed more than one order of magnitude
than the power on the $k/2\pi = 1/10$ pc$^{-1}$ scale,
and obtain the Kolmogorov power spectrum
where the artificial mode conversion due to the $\mathbf{v}=0$ procedure is limited to $\lesssim 5$ percent.

In actual calculations, the shock-compressed layer is significantly deformed 
and is not as simple as our test calculations.
Nevertheless, the physical width of the shock-compressed layer
is $10$--$12$ pc after 1 Myr
and the mean bulk velocity of the shock-compressed layer is limited to $< 1$ km s$^{-1}$, 
at least one order of magnitude smaller than 
the turbulent velocity.
We therefore use this method in our analysis
and employ $k/2\pi=0.1$ pc$^{-1}$
as the proxy of the lowest frequency representing the shock-compressed layer
throughout this article (\eg, when we measure $\chi$ and $b$).

\subsection{The Frequency Range: Sub-Volumes}
\label{subsec:appA2}
In case of the Fourier analysis of the sub-volumes, 
we always set the lowest frequency as the frequency corresponds to
2.56 pc because we analyze the cubic of $($2.56 pc$)^3$.
We here did not apply any spacial care of $\mathbf{v}=0$ as we did above in Section~\ref{subsec:appA1}.
This fixed volume and fixed frequency range
affect the large scatter in the mode ratio among these sub-volumes at early stages 
(Panel (b) of Figure~\ref{fig:PkF_sub}).
Each sub-volume has the different timing 
at which they 
%the sub-volumes 
become completely embedded within the shock-compressed layer.
The layer initially expands almost adiabatically with its width to 2 pc at 0.5 Myr,
then the expansion gradually slows down due to the cooling with the width to 8 pc at 1 Myr,
which corresponds to the separation between the most-separated sub-volumes.
Therefore, it is fair to focus on the evolution of after 1 Myr 
when we discuss the variation among the sub-volumes,
and we report most of our results from 3 Myr in Section~\ref{sec:subvol}.

 %-------------- BIBLIO -------------------------------------------------------

\bibliographystyle{aasjournal}
\bibliography{turbPDF}

 %-----------------------------------------------------------------------------

\end{document}